\newif\ifjcss\jcssfalse    

\ifjcss
\documentclass{elsart}
\else
\documentclass[12pt,twoside]{article}
\usepackage{latexsym}
\fi

\ifjcss 
\def\beq{\begin{equation}}    \def\eeq{\end{equation}}
\def\beqn{\begin{displaymath}}\def\eeqn{\end{displaymath}}
\def\bqa{\begin{eqnarray}}    \def\eqa{\end{eqnarray}}
\def\bqan{\begin{eqnarray*}}  \def\eqan{\end{eqnarray*}}
\else 
\def\,{\mskip 3mu} \def\>{\mskip 4mu plus 2mu minus 4mu} \def\;{\mskip 5mu plus 5mu} \def\!{\mskip-3mu}
\def\dispmuskip{\thinmuskip= 3mu plus 0mu minus 2mu \medmuskip=  4mu plus 2mu minus 2mu \thickmuskip=5mu plus 5mu minus 2mu}
\def\textmuskip{\thinmuskip= 0mu                    \medmuskip=  1mu plus 1mu minus 1mu \thickmuskip=2mu plus 3mu minus 1mu}
\textmuskip
\def\beq{\dispmuskip\begin{equation}}    \def\eeq{\end{equation}\textmuskip}
\def\beqn{\dispmuskip\begin{displaymath}}\def\eeqn{\end{displaymath}\textmuskip}
\def\bqa{\dispmuskip\begin{eqnarray}}    \def\eqa{\end{eqnarray}\textmuskip}
\def\bqan{\dispmuskip\begin{eqnarray*}}  \def\eqan{\end{eqnarray*}\textmuskip}
\fi

\ifjcss
\def\cal{\mathcal}
\else
\newenvironment{keyword}{\centerline{\bf\small
Keyword}\vspace{-1ex}\begin{quote}\small}{\par\end{quote}\vskip
1ex}
\fi

\newtheorem{theorem}{Theorem}
\newtheorem{corollary}[theorem]{Corollary}
\newtheorem{definition}[theorem]{Definition}
\newtheorem{figurex}{Figure}
\def\ftheorem#1#2#3{\begin{theorem}[\boldmath #2]\label{#1} #3 \end{theorem} }
\def\fcorollary#1#2#3{\begin{corollary}[\boldmath #2]\label{#1} #3 \end{corollary} }
\def\fdefinition#1#2#3{\begin{definition}[\boldmath #2]\label{#1} #3 \end{definition} }
\long\def\ffigurex#1#2#3#4{{#4}\begin{figurex}[\boldmath #2]\label{#1}#3\end{figurex}}
\ifjcss
\def\paradot#1{{\itshape{#1.}}}
\def\paranodot#1{{\itshape{#1}}}
\else
\def\myparskip{\vspace{1.5ex plus 0.5ex minus 0.5ex}\noindent}
\def\paradot#1{\myparskip{\bfseries\boldmath{#1.}}}
\def\paranodot#1{\myparskip{\bfseries\boldmath{#1}}}
\fi
\def\toinfty#1{\stackrel{#1\to\infty}{\longrightarrow}}
\def\nq{\hspace{-1em}}
\def\qed{\hspace*{\fill}$\Box\quad$}
\def\odt{{\textstyle{1\over 2}}}
\def\odf{{\textstyle{1\over 4}}}
\def\eps{\varepsilon}                   
\def\epstr{\epsilon}                    
\def\qmbox#1{{\quad\mbox{#1}\quad}}

\def\equa{\stackrel+=}                 
\def\leqa{\stackrel+\leq}              
\def\geqa{\stackrel+\geq}              
\def\eqm{\stackrel\times=}             
\def\leqm{\stackrel\times\leq}
\def\geqm{\stackrel\times\geq}

\def\v#1{{\bf #1}}
\def\l{{\ell}}                          
\def\M{{\cal M}}                        
\def\X{{\cal X}}                        
\def\Y{{\cal Y}}                        
\def\I{{\cal I}}                        
\def\S{{\cal S}}                        
\def\Q{{\cal Q}}
\def\A{{\cal A}}
\def\E{{\bf E}}                         
\def\P{{\bf P}}                         
\def\B{\{0,1\}}                        

\def\MM{M}                              
\def\KM{K\!M}
\def\Km{K\!m}
\def\SetN{{I\!\!N}}  \def\SetR{{I\!\!R}} \def\SetZ{{Z\!\!\!Z}}
\def\lb{\log}
\def\mrcp{the chain rule}
\def\SL{Solomonoff}

\begin{document}

\ifjcss

\begin{frontmatter}
\title{Sequential Predictions based \\ on Algorithmic Complexity}
\author{Marcus Hutter}
\address{IDSIA, Galleria 2, CH-6928 Manno-Lugano, Switzerland \\
marcus@idsia.ch \hspace{9ex} http://www.idsia.ch/$^{_{_\sim}}\!$marcus}

\thanks{Part of this work appeared in the proceedings of
        the 2003 COLT conference \cite{Hutter:03unimdl}}

\else

\title{\vskip -10mm\normalsize\sc Technical Report \hfill IDSIA-16-04
\vskip 2mm\bf\LARGE\hrule height5pt \vskip 5mm
\sc Sequential Predictions based \\ on Algorithmic Complexity
\vskip 6mm \hrule height2pt \vskip 4mm}
\author{{\bf Marcus Hutter}\\[3mm]
\normalsize IDSIA, Galleria 2, CH-6928\ Manno-Lugano, Switzerland\thanks{Part of this work appeared in the proceedings of
               the 2003 COLT conference \cite{Hutter:03unimdl}.}\\
\normalsize marcus@idsia.ch \hspace{8.5ex} http://www.idsia.ch/$^{_{_\sim}}\!$marcus}
\date{\normalsize Submitted: Oct.\ 2003 \hspace{11ex} Published: Oct.\ 2005}
\maketitle

\fi

\begin{abstract}
\noindent This paper studies sequence prediction based on the
monotone Kolmogorov complexity $\Km=-\lb\,m$, i.e.\ based on
universal deterministic/one-part MDL. $m$ is extremely close to
\SL's universal prior $M$, the latter being an excellent predictor
in deterministic as well as probabilistic environments, where
performance is measured in terms of convergence of posteriors or
losses. Despite this closeness to $M$, it is difficult to assess
the prediction quality of $m$, since little is known about the
closeness of their posteriors, which are the important quantities
for prediction. We show that for deterministic computable
environments, the ``posterior'' and losses of $m$ converge, but
rapid convergence could only be shown on-sequence; the
off-sequence convergence can be slow. In probabilistic
environments, neither the posterior nor the losses converge, in
general.
\end{abstract}

\ifjcss\else\def\sep{; }\fi

\ifjcss\end{frontmatter}\else\hfill\fi

\ifjcss\else\pagebreak
\setcounter{tocdepth}{1}
\begin{quote}\begin{quote}
\def\contentsname{\normalsize \hfil Contents \hfil}
{\ifjcss\else\parskip=-2.5ex\fi\boldmath\tableofcontents}
\end{quote}\end{quote}
\fi

\section{Introduction}\label{secIntro}

In this work we study the performance of Occam's razor based
sequence predictors. Given a data sequence $x_1$, $x_2$, ...,
$x_{n-1}$ we want to predict (certain characteristics) of the next
data item $x_n$. Every $x_t$ is an element of some domain $\X$, for
instance weather data or stock-market data at time $t$, or the
$t^{th}$ digit of $\pi$. Occam's razor \cite{Li:97}, appropriately
interpreted, tells us to search for the simplest explanation
(model) of our data $x_1,...,x_{n-1}$ and to use this model for
predicting $x_n$. Simplicity, or more precisely, effective
complexity can be measured by the length of the shortest program
computing sequence $x:=x_1...x_{n-1}$. This length is called the
algorithmic information content of $x$, which we denote by $\tilde
K(x)$. $\tilde K$ stands for one of the many variants of
``Kolmogorov'' complexity (plain, prefix, monotone, ...) or for
$-\lb\,\tilde k(x)$ of universal distributions/measures $\tilde k(x)$.

Algorithmic information theory mainly considers binary sequences.
For finite alphabet $\X$ one could code each $x_t\in\X$ as a
binary string of length $^\lceil\lb|\X|^\rceil$, but this would
not simplify the analysis in this work. The reason being that
binary coding would {\em not} reduce the setting to bit by bit
predictions, but to predict a block of bits before observing the
true block of bits. The only difference in the analysis of general
alphabet versus binary block-prediction is in the convention of
how the length of a string is defined.

The most well-studied complexity regarding its predictive
properties is $\KM(x)=-\lb M(x)$, where $M(x)$ is Solmonoff's
\cite[Eq.(7)]{Solomonoff:64} universal prior. Solomonoff has shown
that the posterior $M(x_t|x_1...x_{t-1})$ rapidly converges to the
true data generating distribution \cite{Solomonoff:78}. In
\cite{Hutter:99errbnd,Hutter:02spupper} it has been shown that $M$
is also an excellent predictor from a decision-theoretic point of
view, where the goal is to minimize loss. In any case, for
prediction, the posterior $M(x_t|x_1...x_{t-1})$, rather than the
prior $M(x_1...x_t)$, is the more important quantity.

Most complexities $\tilde K$ coincide within an additive
logarithmic term, which implies that their ``priors'' $\tilde
k=2^{-\tilde K}$ are close within polynomial accuracy. Some of them
are extremely close to each other.
Many papers deal with the proximity of various complexity measures
\cite[...]{Levin:73random,Gacs:83}. Closeness of two complexity
measures is regarded as indication that the quality of their
prediction is similarly good \cite[p.334]{Li:97}.
On the other hand, besides $M$, little is really known about the
closeness of ``posteriors'', relevant for prediction.

\paradot{Aim and conclusion}
The main aim of this work is to study the predictive properties of
complexity measures other than $\KM$. The monotone complexity
$\Km$ is, in a sense, closest to \SL\ complexity $\KM$. While
$\KM$ is defined via a mixture of infinitely many programs, the
conceptually simpler $\Km$ approximates $\KM$ by the contribution
of the single shortest program. This is also closer to the spirit
of Occam's razor. $\Km$ is a universal deterministic/one-part
version of the popular Minimal Description Length (MDL) principle.
We mainly concentrate on $\Km$ because it has a direct
interpretation as a universal deterministic/one-part MDL
predictor, and it is closest to the excellent performing $\KM$, so
we expect predictions based on other $\tilde K$ not to be better.

The main conclusion we will draw is that closeness of priors does
neither necessarily imply closeness of posteriors, nor good
performance from a decision-theoretic perspective. It is far from
obvious, whether $\Km$ is a good predictor in general, and indeed
we show that $\Km$ can fail (with probability strictly greater
than zero) in the presence of noise, as opposed to $\KM$. We do
not suggest that $\Km$ fails for sequences occurring in practice.
It is not implausible that (from a practical point of view) minor
extra (apart from complexity) assumptions on the environment or
loss function are sufficient to prove good performance of $\Km$.
Some complexity measures like the prefix complexity $K$, fail
completely for prediction.

\paradot{Contents}
{\em Section~\ref{secSetup}} introduces notation and describes how
prediction performance is measured in terms of convergence of
posteriors or losses.
{\em Section~\ref{secMProp}} summarizes known predictive
properties of \SL's prior $M$.
{\em Section~\ref{secASP}} introduces the monotone complexity
$\Km$ and the prefix complexity $K$ and describes how they and
other complexity measures can be used for prediction.
In {\em Section~\ref{secGPF}} we enumerate and relate eight
important properties, which general predictive functions may
posses or not: proximity to $M$, universality, monotonicity, being
a semimeasure, \mrcp, enumerability, convergence, and
self-optimization. Some later needed normalization issues are also
discussed. Furthermore, convergence of non-semimeasures that are
close to $M$ is proven.
{\em Section~\ref{secmProp}} contains our main results. Monotone
complexity $\Km$ is analyzed quantitatively w.r.t.\ the eight
predictive properties. Qualitatively, for deterministic,
computable environments, the posterior converges and is
self-optimizing, but rapid convergence could only be shown
on-sequence; the (for prediction equally important) off-sequence
convergence can be slow. In probabilistic environments, $m$
neither converges, nor is it self-optimizing, in general.
{\em Section~\ref{secFurther}} presents some further results: Poor
predictive performance of the prefix complexity $K$ is shown and a
simpler MDL-inspired way of using $\Km$ for prediction is briefly
discussed.
{\em Section~\ref{secOpen}} contains an outlook and a list of open
question, including the convergence speed of $m$, natural Turing
machines, non-self-optimization for general Turing machines and
losses, other complexity measures, two-part MDL, extra conditions
on environments, and other generalizations.

\section{Notation and Setup}\label{secSetup}

\paradot{Strings and natural numbers}
We write $\X^*$ for the set of finite strings over finite alphabet
$\X$, and $\X^\infty$ for the set of infinity sequences. We use
letters $i,t,n$ for natural numbers, $x,y,z$ for finite strings,
$\epstr$ for the empty string, $\l(x)$ for the length of string
$x$, and $\omega=x_{1:\infty}$ for infinite sequences. We write
$xy$ for the concatenation of string $x$ with $y$. For a string of
length $n$ we write $x_1x_2...x_n$ with $x_t\in\X$ and further
abbreviate $x_{1:n}:=x_1x_2...x_{n-1}x_n$ and $x_{<n}:=x_1...
x_{n-1}$. For a given sequence $x_{1:\infty}$ we say that $x_t$ is
on-sequence and $\bar x_t\neq x_t$ is off-sequence. $x'_t$ may be
on- or off-sequence.

\paradot{Prefix sets/codes}
String $x$ is called a (proper) prefix of $y$ if there is a
$z(\neq\epstr)$ such that $xz=y$. We write $x*=y$ in this case,
where $*$ is a wildcard for a string, and similarly for infinite
sequences. A set of strings is called prefix-free if no element is
a proper prefix of another. A prefix-free set $\cal P$ is also
called a prefix code. Prefix codes have the important property of
satisfying Kraft's inequality $\sum_{x\in\cal P} |\X|^{-\l(x)}\leq
1$.

\paradot{Asymptotic notation}
We abbreviate $\lim_{t\to\infty}[f(t)-g(t)]=0$ by
$f(t)\toinfty{t}g(t)$ and say $f$ converges to $g$, without
implying that $\lim_{t\to\infty}g(t)$ itself exists.
The big $O$-notation $f(x)=O(g(x))$ means that there are constants
$c$ and $x_0>0$ such that $|f(x)|\leq c|g(x)|\,\forall x>x_0$. The
small $o$-notation $f(x)=o(g(x))$ abbreviates
$\lim_{x\to\infty}f(x)/g(x)=0$.
We write $f(x)\leqm  g(x)$ for $f(x)=O(g(x))$ and
$f(x)\leqa  g(x)$ for $f(x)\leq g(x)+O(1)$.
Corresponding equalities can be defined similarly. They hold if
the corresponding inequalities hold in both directions.
$\sum_{t=1}^\infty a_t^2<\infty$ implies $a_t\toinfty{t} 0$. We say
that $a_t$ converges fast or rapidly to zero if
$\sum_{t=1}^\infty a_t^2\leq c$, where $c$ is a constant of
reasonable size; $c=100$ is reasonable, maybe even $c=2^{30}$, but
$c=2^{500}$ is not.$\!$\footnote{Environments
of interest have reasonable complexity $K$,
but $2^K$ is not of reasonable size.}
The number of times for which $a_t$ deviates
from 0 by more than $\eps$ is finite and bounded by $c/\eps^2$;
no statement is possible for {\em which} $t$ these deviations occur.
The cardinality of a set $\cal S$ is denoted by $|{\cal S}|$ or
$\#\cal S$.
For properties $A(t)\in\{true,false\}$ we say
\begin{center}
\begin{tabular}{c||c|c|c|c}
  $A(t)$ is valid for ... $t$          & almost all  & most         & many       & finitely many \\\hline
  {\em iff} $\;\;\#\{t\leq n:A(t)\}$   & $\equa\; n$ & $=\; n-o(n)$ & $\eqm\; n$ & $\leq\; c\quad (\exists c)$
\end{tabular}
\end{center}

\paradot{(Semi)measures}
We call $\rho:\X^*\to[0,1]$ a (semi)measure {\em iff}
$\sum_{x_n\in\X}\rho(x_{1:n}) \stackrel{(<)}= \rho(x_{<n})$ and
$\rho(\epstr) \stackrel{(<)}= 1$.
$\rho(x)$ is interpreted as the $\rho$-probability of sampling a
sequence which starts with $x$.
In case of a semimeasure the gap
$g_n=1-\sum_{x_{1:n}}\rho(x_{1:n})\geq 0$ may be interpreted as the
possibility/probability of {\em finite} sequences of length less
than $n$ \cite{Zvonkin:70,Schmidhuber:00toe}, or as an evidence gap in
Dempster-Shafer theory \cite{Dempster:68,Shafer:76}.
The conditional probability (posterior)
\beq\label{defBayes}
  \rho(x_t|x_{<t}):={\rho(x_{1:t})\over\rho(x_{<t})}
\eeq
is the $\rho$-probability that a string $x_1...x_{t-1}$ is
followed by (continued with) $x_t$. We call $\rho$ deterministic
if $\exists\omega:\rho(\omega_{1:n})=1$ $\forall n$. In this case
we identify $\rho$ with $\omega$.

\paradot{Convergent predictors}
We assume that $\mu$ is the ``true''\footnote{Also called {\em
objective} or {\em aleatory} probability or {\em chance}.}
sequence generating measure, also called environment. If we know
the generating process $\mu$, and given past data $x_{<t}$ we can
predict the probability $\mu(x_t|x_{<t})$ of the next data item
$x_t$. Usually we do not know $\mu$, but estimate it from
$x_{<t}$. Let $\rho(x_t|x_{<t})$ be an estimated
probability\footnote{Also called {\em subjective} or {\em belief}
or {\em epistemic} probability.} of $x_t$, given $x_{<t}$.
Closeness of $\rho(x_t|x_{<t})$ to $\mu(x_t|x_{<t})$ is expected
to lead to ``good'' predictions:

Consider, for instance, a weather data sequence $x_{1:n}$ with
$x_t=1$ meaning rain and $x_t=0$ meaning sun at day $t$. Given
$x_{<t}$ the probability of rain tomorrow is $\mu(1|x_{<t})$. A
weather forecaster may announce the probability of rain to be
$y_t:=\rho(1|x_{<t})$, which should be close to
the true probability $\mu(1|x_{<t})$.
To aim for
\beq\label{eqconv}
 \rho(x'_t|x_{<t}) \;\stackrel{(fast)}\longrightarrow\; \mu(x'_t|x_{<t})
 \qmbox{for} t\to\infty
\eeq
seems reasonable. A sequence of random variables $z_t=z_t(\omega)$
(like $z_t=\rho(x_t|x_{<t})-\mu(x_t|x_{<t})$) is said to converge
to zero with $\mu$-probability 1 (w.p.1) if the set $\{\omega :
z_t(\omega)\toinfty{t} 0\}$ has $\mu$-measure 1. $z_t$ is said to
converge to zero in mean sum (i.m.s) if
$\sum_{t=1}^\infty\E[z_t^2]\leq c<\infty$, where $\E$ denotes
$\mu$-expectation. Convergence i.m.s.\ implies convergence w.p.1
(rapid if $c$ is of reasonable size).

Depending on the interpretation, a $\rho$ satisfying
(\ref{eqconv}) could be called consistent or self-tuning
\cite{Kumar:86}. One problem with using (\ref{eqconv}) as
performance measure is that closeness cannot be computed, since
$\mu$ is unknown. Another disadvantage is that (\ref{eqconv}) does
not take into account the value of correct predictions or the
severity of wrong predictions.

\paradot{Self-optimizing predictors}
More practical and flexible is a decision-theoretic approach,
where performance is measured w.r.t.\ the true outcome sequence
$x_{1:n}$ by means of a loss function, for instance
$\ell_{x_ty_t}:=(x_t-y_t)^2$, which does not involve $\mu$.
More generally, let $\ell_{x_t y_t}\in[0,1]\subset\SetR$ be the
received loss when performing some prediction/decision/action
$y_t\in\Y$ and $x_t\in\X$ is the $t^{th}$ symbol of the sequence.
Let $y_t^\Lambda \in \Y$ be the prediction of a (causal)
prediction scheme $\Lambda$. The true probability of the next
symbol being $x_t$, given $x_{<t}$, is $\mu(x_t|x_{<t})$. The
$\mu$-expected loss (given $x_{<t}$) when $\Lambda$ predicts the
$t^{th}$ symbol is
\beqn\label{loss2}
  l_t^\Lambda(x_{<t}) \;:=\;
  \sum_{x_t}\mu(x_t|x_{<t}) \ell_{x_t y_t^\Lambda}.
\eeqn
The goal is
to minimize the $\mu$-expected loss. More generally, we define the
$\Lambda_\rho$ sequence prediction scheme
\beq\label{xlrdef}
  y_t^{\smash{\Lambda_\rho}} :=
  \arg\min_{y_t\in\Y}\sum_{x_t}\rho(x_t|x_{<t})\ell_{x_t y_t},
\eeq
which minimizes the $\rho$-expected loss. If $\mu$ is known,
$\Lambda_\mu$ is obviously the best prediction scheme in the sense
of achieving minimal expected loss ($l_t^{\smash{\Lambda_\mu}}\leq
l_t^\Lambda$ for all $\Lambda$). An important special case is
the error loss $\ell_{xy}=1-\delta_{xy}$ with $\Y=\X$. In this
case $\Lambda_\rho$ predicts the $y_t$ which maximizes
$\rho(y_t|x_{<t})$,  and $\sum_t\E[l_t^{\smash{\Lambda_\rho}}]$ is the
expected number of prediction errors (where
$y_t^{\smash{\Lambda_\rho}}\neq x_t$).
The natural decision-theoretic counterpart of (\ref{eqconv}) is to
aim for
\beq\label{eqlconv}
 l_t^{\smash{\Lambda_\rho}}(x_{<t}) \;\stackrel{(fast)}\longrightarrow\;
 l_t^{\smash{\Lambda_\mu}}(x_{<t}) \qmbox{for} t\to\infty
\eeq
what is called (without the fast supplement) self-optimization
in control-theory \cite{Kumar:86}.

\section{Predictive Properties of $M=2^{-\KM}$}\label{secMProp}

We define a prefix/monotone Turing machine $T$ as a Turing machine
with a binary unidirectional input tape, an unidirectional output
tape with alphabet $\X$, and some bidirectional work tapes.
We say $T$ halts on input $p$ with output $x$ and write ``$T(p)=x$
halts'' if $p$ is to the left of the input head and $x$ is to the
left of the output head after $T$ halts. The set of $p$ on which
$T$ halts forms a prefix code. We call such codes $p$ {\em
self-delimiting} programs.
We write $T(p)=x*$ if $T$ outputs a string starting with $x$; $T$
need not to halt in this case. $p$ is called {\em minimal} if
$T(q)\neq x*$ for all proper prefixes of $p$.
The set of all prefix/monotone Turing machines $\{T_1,T_2,...\}$ can
be effectively enumerated.
There exists a universal prefix/monotone Turing machine $U$ which
can simulate every $T_i$.
A function is called computable if there is a Turing machine
which computes it. A function is called enumerable if it can be
approximated from below. Let $\M_{comp}^{msr}$ be the set of all
computable measures, $\M_{enum}^{semi}$ the set of all enumerable
semimeasures, and $\M_{det}$ be the set of all deterministic
measures ($\widehat=\X^\infty$).$\!$\footnote{$\M_{enum}^{semi}$
is enumerable, but $\M_{comp}^{msr}$ is not, and $\M_{det}$ is
uncountable.}

Levin \cite{Zvonkin:70,Li:97} has shown the existence of an
enumerable universal semimeasure $\MM$ ($\MM\geqm\nu$
$\forall\nu\in\M_{enum}^{semi}$). An explicit expression due to
Solomonoff \cite[Eq.(7)]{Solomonoff:64} is
\beq\label{defM}\label{defKM}
  \MM(x) \;:=\; \sum_{p:U(p)=x*}2^{-\l(p)}, \qquad
  \KM(x) := -\lb M(x).
\eeq
The sum is over all (possibly nonhalting) minimal programs $p$
which output a string starting with $x$. This definition is
equivalent to the probability that $U$ outputs a string starting
with $x$ if provided with fair coin flips on the input tape. $M$
can be used to characterize randomness of individual sequences: A
sequence $x_{1:\infty}$ is (Martin-L\"{o}f) $\mu$-random, {\em
iff} $\exists c:M(x_{1:n})\leq c\cdot\mu(x_{1:n})\forall n$.
For later comparison, we summarize the (excellent) predictive
properties of $\MM$
\cite{Solomonoff:78,Hutter:01alpha,Hutter:02spupper,Hutter:04uaibook}
(the numbering will become clearer later):

\ftheorem{thMProp}{Properties of $\MM=2^{-\KM}$}{
\SL's prior $\MM$ defined in (\ref{defM}) is a $(i)$
universal, $(v)$ enumerable, $(ii)$ monotone, $(iii)$ semimeasure,
which $(vi)$ converges to $\mu$ i.m.s., and $(vii)$ is
self-optimizing i.m.s. More quantitatively:
\begin{list}{}{\parsep=1ex\itemsep=0ex}
\item[$(vi)$]
  $\sum_{t=1}^\infty\E[\sum_{x'_t}(M(x'_t|x_{<t})-\mu(x'_t|x_{<t}))^2]
  \;\leqa\; \ln 2\cdot K(\mu)$, which implies \\
  $M(x'_t|x_{<t}) \;\toinfty{t}\; \mu(x'_t|x_{<t})$ i.m.s.\ for
  $\mu\in\M_{comp}^{msr}$.
\item[$(vii)$]
  $\sum_{t=1}^\infty\E[(l_t^{\smash{\Lambda_M}}-l_t^{\smash{\Lambda_\mu}})^2]
  \;\leqa\;  2\ln 2\cdot K(\mu)$, which implies \\
  $l_t^{\smash{\Lambda_M}} \;\toinfty{t}\; l_t^{\smash{\Lambda_\mu}}$ i.m.s.\ for
  $\mu\in\M_{comp}^{msr}$,
\end{list}
where $K(\mu)$ is the length of the shortest program computing
function $\mu$.
}

\section{Alternatives to \SL's Prior $\MM$}\label{secASP}

The goal of this work is to investigate whether some other
quantities that are closely related to $\MM$ also lead to good
predictors. The prefix Kolmogorov complexity $K$ is closely
related to $\KM$ ($K(x)=\KM(x)+O(\log\,\l(x))$). $K(x)$ is defined
as the length of the shortest halting program on $U$ with output
$x$:
\beq\label{defK}\label{defk}
  K(x) := \min\{\l(p): U(p)=x \mbox{ halts} \}, \qquad
  k(x):=2^{-K(x)}.
\eeq
In Section \ref{secPPK} we briefly discuss that $K$ completely
fails for predictive purposes. More promising is to approximate
$\MM(x)=\sum_{p:U(p)=x*}2^{-\l(p)}$ by the dominant contribution
in the sum, which is given by
\beq\label{defm}\label{defKm}
  m(x):=2^{-\Km(x)} \qmbox{with}
  \Km(x):=\min_p\{\l(p):U(p)=x*\}.
\eeq
$\Km$ is called {\em monotone complexity} and has been shown to be
{\em very} close to $\KM$ \cite{Levin:73random,Gacs:83} (see
Theorem \ref{thmDProp}$(o)$). It is natural to call a sequence
$x_{1:\infty}$ {\em computable} if $\Km(x_{1:\infty})<\infty$.
$\KM$, $\Km$, and $K$ are ordered in the following way:
\beq\label{Krels}
\ifjcss\nq
  0 \leq K(x|\l(x)) \leqa \KM(x) \leq \Km(x) \leq K(x)
    \leqa \l(x)\!\cdot\!\lb|\X| \!+\! 2\lb\l(x).
\else
  0 \;\leq\; K(x|\l(x)) \;\leqa\; \KM(x) \;\leq\; \Km(x) \;\leq\; K(x)
  \;\leqa\; \l(x)\!\cdot\!\lb|\X| + 2\lb\l(x).
\fi
\eeq
The second inequality follows from the fact that, given $n$
and Kraft's inequality $\sum_{x\in\X^n}M(x)\leq 1$,
there exists for $x\in\X^n$ a Shannon-Fano code of length
$-\lb M(x)$, which is effective since $M$ is enumerable.
The other inequalities are obvious from the definitions.
There are many complexity measures (prefix, \SL, monotone,
plain, process, extension, ...) which we generically denote by
$\tilde K\in\{K,\KM,\Km,...\}$ and their associated ``predictive
functions'' $\tilde k(x):=2^{-\tilde K(x)}\in\{k,M,m,...\}$. This
work is mainly devoted to the study of $m$.

Note that $\tilde k$ is generally not a semimeasure, so we have to
clarify what it means to predict using $\tilde k$. One popular
approach which is at the heart of the (one-part) MDL principle is
to predict the $y$ which minimizes $\tilde K(xy)$ (maximizes
$\tilde k(xy))$, where $x$ are past given data:
$y_t^{MDL}:=\arg\min_{y_t}\tilde K(x_{<t}y_t)$.

For complexity measures $\tilde K$, the conditional version
$\tilde K_|(x|y)$ is often defined\footnote{Usually written
without index $|$.} as $\tilde K(x)$, but where the underlying
Turing machine $U$ has additionally access to $y$.
The definition $\tilde k_|(x|y):=2^{-\tilde K_|(x|y)}$ for the
conditional predictive function $\tilde k$ seems natural, but has
the disadvantage that the crucial chain rule (\ref{defBayes}) is
violated. For $\tilde K=K$ and $\tilde K=\Km$ and most other
versions of $\tilde K$, \mrcp\ is still satisfied
approximately (to logarithmic accuracy), but this is not
sufficient to prove convergence (\ref{eqconv}) or
self-optimization (\ref{eqlconv}). Therefore, we define $\tilde
k(x_t|x_{<t}):=\tilde k(x_{1:t})/\tilde k(x_{<t})$ in the
following, analogously to semimeasures $\rho$ (like $\MM$). A
potential disadvantage of this definition is that $\tilde
k(x_t|x_{<t})$ is not enumerable, whereas $\tilde k_|(x_t|x_{<t})$
and $\tilde k(x_{1:t})$ are.

We can now embed MDL predictions minimizing $\tilde K$ into our
general framework: MDL coincides with the $\Lambda_{\tilde k}$
predictor for the error loss:
\beq\label{eqMDLk}
\ifjcss\nq\!
  y_t^{\smash{\Lambda_{\tilde k}}} =
  \arg\max_{y_t}\tilde k(y_t|x_{<t}) =
  \arg\max_{y_t}\tilde k(x_{<t}y_t) =
  \arg\min_{y_t}\tilde K(x_{<t}y_t) =
  y_t^{MDL}
\else
  y_t^{\smash{\Lambda_{\tilde k}}} \;=\;
  \arg\max_{y_t}\tilde k(y_t|x_{<t}) \;=\;
  \arg\max_{y_t}\tilde k(x_{<t}y_t) \;=\;
  \arg\min_{y_t}\tilde K(x_{<t}y_t) \;=\;
  y_t^{MDL}
\fi
\eeq
In the first equality we inserted $\ell_{xy}=1-\delta_{xy}$ into
(\ref{xlrdef}). In the second equality we used \mrcp\
(\ref{defBayes}). In both steps we dropped some in $\arg\max$
ineffective additive/multiplicative terms independent of $y_t$. In
the third equality we used $\tilde k=2^{-\tilde K}$. The last
equality formalizes the one-part MDL principle: given $x_{<t}$
predict the $y_t\in\X$ which leads to the shortest code $p$.
Hence, validity of (\ref{eqlconv}) tells us something about the
validity of the MDL principle. (\ref{eqconv}) and (\ref{eqlconv})
address what (good) prediction {\em means}.

\section{General Predictive Functions}\label{secGPF}

We have seen that there are predictors (actually the major one
studied in this work) $\Lambda_\rho$, but where $\rho(x_t|x_{<t})$
is not (immediately) a semimeasure. Nothing prevents us from
replacing $\rho$ in (\ref{xlrdef}) by an arbitrary function
$b_|:\X^*\to[0,\infty)$, written as $b_|(x_t|x_{<t})$.
We also define general functions $b:\X^*\to[0,\infty)$, written as
$b(x_{1:n})$ and $b(x_t|x_{<t}):={b(x_{1:t})\over b(x_{<t})}$,
which may not coincide with $b_|(x_t|x_{<t})$. Most terminology for
semimeasure $\rho$ can and will be carried over to the case of
general predictive functions $b$ and $b_|$, but one has to be
careful which properties and interpretations still hold:

\fdefinition{defbProp}{Properties of predictive functions}{
We call functions $b,b_|:\X^*\to[0,\infty)$ (conditional)
predictive functions. They may possess some of the following
properties:
\begin{list}{}{\parsep=1ex\itemsep=0ex\leftmargin=5ex\labelwidth=5ex}
\item[$o)$] {\em Proximity:} $b(x)$ is ``close'' to the universal
prior $\MM(x)$
\item[$i)$] {\em Universality:} $b\geqm\M$, i.e.\
$\forall\nu\in\M\,\exists c>0: b(x)\geq c\cdot\nu(x)\forall x$.
\item[$ii)$] {\em Monotonicity:} $b(x_{1:t})\leq
b(x_{<t})\;\forall t,x_{1:t}$
\item[$iii)$] {\em Semimeasure:} $\sum_{x_t}b(x_{1:t})\leq
b(x_{<t})$ and $b(\epstr)\leq 1$
\item[$iv)$] {\em Chain rule:} $b(x_{1:t})=b.(x_t|x_{<t})b(x_{<t})$
\item[$v)$] {\em Enumerability:} $b$ is lower semicomputable
\item[$vi)$] {\em Convergence:}
$b.(x'_t|x_{<t})\toinfty{t}\mu(x'_t|x_{<t})$ $\forall \mu\in\M,
x'_t\in\X$ i.m.s.\ or w.p.1
\item[$vii)$] {\em Self-optimization:}
$l_t^{\smash{\Lambda_{b.}}} \;\toinfty{t}\; l_t^{\smash{\Lambda_\mu}}$ i.m.s.\ or w.p.1
\end{list}
where $b.$ refers to $b$ or $b_|$
}

\noindent The importance of the properties $(i)-(iv)$ stems from
the fact that they together imply convergence $(vi)$ and
self-optimization $(vii)$. Regarding proximity $(o)$ we left
open what we mean by ``close''. We also did not specify $\M$ but
have in mind all computable measures $\M_{comp}^{msr}$ or
enumerable semimeasures $\M_{enum}^{semi}$, possibly restricted to
deterministic environments $\M_{det}$.

\ftheorem{thPredRel}{Predictive relations}{\hfill
\begin{list}{}{\parsep=1ex\itemsep=0ex\leftmargin=5ex\labelwidth=5ex}
\item[$a)$] $(iii)\Rightarrow(ii)$: A semimeasure is monotone.
\item[$b)$] $(i),(iii),(iv)\Rightarrow(vi)$: The posterior $b.$ as defined
by \mrcp\ $(iv)$ of a universal semimeasure $b$ converges to
$\mu$ i.m.s.\ for all $\mu\in\M$.
\item[$c)$] $(i),(iii),(v)\Rightarrow(o)$: Every w.r.t.\ $\M_{enum}^{semi}$
universal enumerable semimeasure coincides with $M$ within a multiplicative constant.
\item[$d)$] $(vi)\Rightarrow(vii)$: Posterior convergence i.m.s./w.p.1
implies self-optimization i.m.s./w.p.1.
\end{list}
}

\paradot{Proof sketch} %
$(a)$ follows trivially from dropping the sum in $(iii)$, %
$(b)$ is Solomonoff's major result
\cite{Solomonoff:78,Li:97,Hutter:01alpha,Hutter:04uaibook}, %
$(c)$ is due to Levin \cite{Zvonkin:70}, %
$(d)$ follows from $0\leq l_t^{\smash{\Lambda_{b.}}}-l_t^{\smash{\Lambda_\mu}}
\leq\sum_{x'_t}|b.(x'_t|x_{<t})-\mu(x'_t|x_{<t})|$, since
$\ell\in[0,1]$ \cite[Thm.4$(ii)$]{Hutter:02spupper}.
\qed

We will see that $(i),(iii),(iv)$ are crucial for proving
$(vi),(vii)$.

\paradot{Normalization}
Let us consider a scaled $b$ version
$b_{norm}(x_t|x_{<t}):=c(x_{<t})b(x_t|x_{<t})$, where $c>0$ is
independent of $x_t$. Such a scaling does not affect the
prediction scheme $\Lambda_b$ (\ref{xlrdef}), i.e.\
$y_t^{\smash{\Lambda_b}}=y_t^{\smash{\Lambda_{b_{norm}}}}$, which implies
$l_t^{\smash{\Lambda_{b_{norm}}}}=l_t^{\smash{\Lambda_b}}$. Convergence
$b(x'_t|x_{<t})\to\mu(x'_t|x_{<t})$ implies
$\sum_{x'_t}b(x'_t|x_{<t})\to 1$ if $\mu$ is a measure, hence also
$b_{norm}(x'_t|x_{<t})\to\mu(x'_t|x_{<t})$
for\footnote{Arbitrarily we define
$b_{norm}(x_t|x_{<t})={1\over|\X|}$ if
$\sum_{x'_t}b(x'_t|x_{<t})=0$.}
$c(x_{<t}):=[\sum_{x'_t}b(x'_t|x_{<t})]^{-1}$. Speed of
convergence may be affected by normalization, either positively or
negatively. Assuming \mrcp\ (\ref{defBayes}) for $b_{norm}$ we get
\beqn
  \ifjcss\nq\!\!\fi
  b_{norm}(x_{1:n}) =
  \prod_{t=1}^n{b(x_{1:t})\over\sum_{x_t}b(x_{1:t})} =
  d(x_{<n})b(x_{1:n}),
  \ifjcss\quad\else\qquad\fi
  d(x_{<n}):={1\over b(\epstr)}\prod_{t=1}^n{b(x_{<t})\over\sum_{x_t}b(x_{1:t})}
\eeqn
Whatever $b$ we start with, $b_{norm}$ is a measure, i.e.\ $(iii)$
is satisfied with equality. Convergence and self-optimization
proofs are now eligible for $b_{norm}$, provided universality $(i)$
can be proven for $b_{norm}$.
If $b$ is a semimeasure, then $d\geq 1$, hence
$\MM_{norm}\geq\MM\geqm\M_{enum}^{semi}$ is universal and
converges $(vi)$ with the same bound (Theorem \ref{thMProp}$(vi)$)
as for $\MM$. On the other hand, $d(x_{<n})$ may be unbounded for
$b=k$ and $b=m$, so normalization does not help us in these cases
for proving $(vi)$. Normalization transforms a universal
non-semimeasure into a measure, which may no longer be universal.

\paradot{Universal Non-Semimeasures}\label{secUNSM}
If $b\geqm\MM$ is a universal semimeasure, then $b$ is as good for
prediction as $\MM$. The bounds are loosened by at most an
additive constant. For $b$ still dominating $\MM$, but no longer
being a semimeasure, we believe that $(vi)$ and $(vii)$ can be
violated. Bounds can be shown without any further assumptions on $b$
on-sequence and if we demand a lower {\em and} upper bound on $b$,
i.e.\ $b\eqm  M$, then also off-sequence:

\ftheorem{thCUNSM}{Convergence of Universal Non-Semimeasures}{
For every predictive function $b$, and real numbers $a$ and $c$ it holds:
\beqn\ifjcss\nq\nq\fi
\begin{array}{crcl}
  a) & \sum_{t=1}^n 1-b(x_t|x_{<t}) & \leq & \ln 2\cdot\KM(x_{1:n})+\ln a^{-1} \qmbox{if} a\MM(x)\leq b(x)\,\forall x, \\
  b) & \sum_{t=1}^n\sum_{\bar x_t\neq x_t} b(\bar x_t|x_{<t}) & \leq & {c\over a}\ln 2\cdot \KM(x_{1:n}) \qmbox{if} a\MM(x)\leq b(x)\leq c\MM(x)\,\forall x. \\
\end{array}
\eeqn
For computable $x_{1:\infty}$ this implies: $b(\bar x_t|x_{<t})\to
0$ and $b_{norm}(\bar x_t|x_{<t})\to 0$ for $\bar x_t\neq x_t$, and
$b(x_t|x_{<t})\to 1$ if $b(x_t|x_{<t})\leq 1$ and
$b_{norm}(x_t|x_{<t})\to 1$ for $t\to\infty$.
}

\paradot{Remarks} If $b$ additionally is a semimeasure,
i.e.\ $\sum_{\bar x_t\neq x_t} b(\bar x_t|x_{<t})\leq
1-b(x_t|x_{<t})$ then (a) implies an improved off-sequence bound.
Note that $b(\bar x_t|x_{<t})\to 0$ does not imply
$b(x_t|x_{<t})\to 1$. Furthermore, although $b_{norm}$ is a
measure, convergence cannot be concluded similarly to
(\ref{eqCUSM}), since $b_{norm}$ may not be universal due to a
possibly unbounded normalizer $d(x_{<t})$.

\paradot{Proof}\vspace{-3ex}
\bqan\label{eqCUSM}
  \nq\nq{\bf(a)}\qquad \sum_{t=1}^n 1- b(x_t|x_{<t})
  &\;\leq\;& \sum_{t=1}^n \ln b(x_t|x_{<t})^{-1}
  \;=\; \ln b(x_{1:n})^{-1}
\\
  &\;\leq\;& \ln [a M(x_{1:n})]^{-1}
  \;=\; \ln 2\cdot \KM(x_{1:n})+\ln a^{-1}
\eqan
$\bf (b)$
\beqn\ifjcss\nq\!\!\!\!\fi
  b(\bar x_t|x_{<t})
  \;\leq\; b(\bar x_t|x_{<t}) \!\cdot\! {b(x_{<t})\over a M(x_{<t})}
  = {b(x_{<t}\bar x_t)\over a M(x_{<t})}
  \;\leq\; {c M(x_{<t}\bar x_t)\over a M(x_{<t})}
  \;=\; {c\over a} M(\bar x_t|x_{<t}).
\eeqn
For every semimeasure it holds:
\beqn
  \sum_{t=1}^n\sum_{\bar x_t\neq x_t}\!\!\rho(\bar x_t|x_{<t})
  \;\leq\; \sum_{t=1}^n 1\!-\!\rho(x_t|x_{<t})
  \;\leq\; -\! \sum_{t=1}^n \ln\rho(x_t|x_{<t})
  \;=\; - \ln\rho(x_{1:n})
\eeqn
Combining both bounds and using that $M$ is a semimeasure we get
\beqn
  \sum_{t=1}^n\sum_{\bar x_t\neq x_t} b(\bar x_t|x_{<t})
  \;\leq\; {c\over a}\sum_{t=1}^n\sum_{\bar x_t\neq x_t} M(\bar x_t|x_{<t})
  \;\leq\; {c\over a}\ln 2 \!\cdot\! \KM(x_{1:n}).
\eeqn\vspace{-2ex}
\qed

\section{Predictive Properties of $m=2^{-\Km}$}\label{secmProp}

We can now state which predictive properties of $m$ hold, and
which not. We first summarize
the qualitative predictive properties of $m$ in Corollary
\ref{thmProp}, and subsequently present detailed quantitative
results in Theorems~\ref{thmDProp}$(o)-(vii)$, followed by an
item-by-item explanation, discussion and detailed proofs.

\fcorollary{thmProp}{Properties of $m=2^{-\Km}$}{
For $b=m=2^{-\Km}$, where $\Km$ is the monotone Kolmogorov
complexity (\ref{defKm}), the following properties of Definition
\ref{defbProp} are satis\-fied\-/\-vio\-lated: %
$(o)$ For every $\mu\in\M_{comp}^{msr}$ and every $\mu$-random
sequence $x_{1:\infty}$, $m(x_{1:n})$ equals $M(x_{1:n})$ within a
multiplicative constant. %
$m$ is $(i)$ universal (w.r.t.\ $\M=\M_{comp}^{msr}$), %
$(ii)$ monotone, %
and $(v)$ enumerable, %
but is $\neg(iii)$ not a semimeasure. %
$m$ satisfies $(iv)$ \mrcp\ by definition for
$m.=m$, but for $m.=m_|$ \mrcp\ is only satisfied to logarithmic order. %
For $m.=m$, $m$ $(vi)$ converges and $(vii)$ is self-optimizing
for deterministic $\mu\in\M_{comp}^{msr}\cap\M_{det}$, but in
general not for probabilistic $\mu\in\M_{comp}^{msr}\setminus\M_{det}$. %
}

\noindent The lesson to learn is that although $m$ is very close
to $M$ in the sense of $(o)$ and $m$ dominates all computable
measures $\mu$, predictions based on $m$ may nevertheless fail
(cf. Theorem~\ref{thMProp}).

\paradot{Some proof ideas}
$(o)$ \cite[Thm.3.4]{Zvonkin:70} and \cite{Levin:73random}. %
$(i)$ \cite{Levin:73random}.
$(ii)$ from $\Km(xy)\geq \Km(x)$ (see definition of $\Km$). %
$\neg(iii)$ follows from $(i),(iv),\neg(vi)$ and Theorem
\ref{thPredRel}$b$ with $m_|:=m$. %
$(iv)$ follows within $\log$ from $\Km=K+O(\log)$ and
\cite[Thm.3.9.1]{Li:97}, %
$\neg(iv)$, since it does not even hold within
an additive constant. %
$(v)$ immediate from definition. %
$(vi)$ similarly as for $\MM$. %
$\neg(vi)$ Use $m_|\in 2^{-\SetN\!_0}$ and define a $\mu_|\not\in
2^{-\SetN\!_0}$. %
$(vii)$ follows from $(vi)$. %
$\neg(vii)$ For the monotone Turing machine $U$ defined by
$U(1x0)=x0$, the loss $\ell_{00}=\ell_{11}=0$, $\ell_{10}=1$,
$\ell_{01}={2\over 3}$ and a Bernoulli($\odt$) process
$\mu(x_t|x_{<t})=\odt$ one can show $y_t^{\smash{\Lambda_m}}=0\neq
1=y_t^{\smash{\Lambda_\mu}}$, which implies
$l_t^{\smash{\Lambda_m}}=\odt>{1\over
3}=l_t^{\smash{\Lambda_\mu}}$. Extending $U$ to a universal Turing
machine by $U(0^{s+1}p)=U'(p)$ leaves this result intact with
probability $\geq 1-2^{-s}$, since random strings cannot be
compressed (by $U'$). \qed

\setcounter{subsection}{-1}
\subsection{Proximity of $m=2^{-\Km}$}

The following closeness/separation results between $\Km$ and $\KM$
are known:

\ftheorem{thmo}{o) (Proximity of $m=2^{-\Km}$}{\hfill
\begin{list}{}{\parsep=1ex\itemsep=0ex\leftmargin=5ex\labelwidth=5ex}
\item[$\scriptstyle(1)$]
$\forall\mu\in\M_{comp}^{msr}\,\forall\mu$-random $\omega\,\exists
c_\omega \,:\, \Km(\omega_{1:n})\leq \KM(\omega_{1:n})+c_\omega\,\forall
n$, \hfill \cite{Levin:73random}
\item[$\scriptstyle(2)$]
$\KM(x)\leq \Km(x)\leq \KM(x)+2\,\lb \Km(x)+O(1)\,\forall x$. \hfill \cite[Thm.3.4]{Zvonkin:70}
\item[$\scriptstyle\neg(3)$]
$\forall c \,:\, \Km(x)-\KM(x)\geq c$ for infinitely many $x$. \hfill \cite{Gacs:83}
\end{list}
}

\paradot{Remarks}
The first line $(o_1)$ shows that $m$ is close to $M$
within a multiplicative constant for nearly all strings in a very
strong sense. $\sup_n{M(\omega_{1:n})\over m(\omega_{1:n})}\leq
2^{c_\omega}$ is finite for every $\omega$ which is random (in
the sense of Martin-L{\"o}f) w.r.t.\ {\em any} computable $\mu$,
but note that the constant $c_\omega$ depends on $\omega$.
Levin falsely conjectured the result to be true for {\em all}
$\omega$, but could only prove it to hold within logarithmic
accuracy $(o_2)$.
A later result by G\'acs $\neg(o_3)$, indeed, shows that
$\Km-\KM$ is unbounded (for infinite alphabet it can even increase
logarithmically).

\paradot{Proof}
The first two properties are due to Levin and are proven in
\cite{Levin:73random} and \cite[Thm.3.4]{Zvonkin:70}, respectively.
The third property follows easily from G\'acs result
\cite{Gacs:83}, which says that if $g$ is some monotone
co-enumerable function for which $\Km(x)-\KM(x)\leq g(\l(x))$
holds for all $x$, then $g(n)$ must be $\geqa  K(n)$. Assume
$\Km(x)-\KM(x)\geq \lb\,\l(x)$ only for finitely many $x$.
Then there exists a $c$ such that $\Km(x)-\KM(x)\leq \lb\,\l(x)+c$
for {\em all} $x$. G\'acs' theorem now implies $\lb\,n+c\geqa
K(n)\,\forall n$, which is wrong due to Kraft's inequality $\sum_n
2^{-K(n)}\leq 1$.
\qed

\subsection{Universality of $m=2^{-\Km}$}

\addtocounter{theorem}{-1}
\ftheorem{thmi}{i) (Universality of $m=2^{-\Km}$}{\hfill
\begin{list}{}{\parsep=1ex\itemsep=0ex\leftmargin=5ex\labelwidth=5ex}
\item[$\scriptstyle(1)$]
$\Km(x)\leqa -\lb\,\mu(x)+K(\mu) \qmbox{if}
 \mu\in\M_{comp}^{msr}$, \hfill \cite[Thm.4.5.4]{Li:97}
\item[$\scriptstyle(2)$]
$m\geqm\M_{comp}^{msr}, \qmbox{but}
 m\not\geqm\M_{enum}^{semi}$ (unlike
$\MM\geqm\M_{enum}^{semi}$).
\end{list}
}

\paradot{Remarks} The first line $(i_1)$ can be interpreted as a
``continuous'' coding theorem for $\Km$ and recursive $\mu$. It
implies (by exponentiation) that $m$ dominates all computable
measures $(i_2)$. Unlike $\MM$ it does {\em not} dominate all
enumerable semimeasures. Dominance is a key feature for good
predictors. From a practical point of view the assumption that the
true generating distribution $\mu$ is a proper measure and
computable seems not to be restrictive. The problem will be that
$m$ is not a semimeasure.

\paradot{Proof}
The first line is proven in \cite[Thm.4.5.4]{Li:97}. Exponentiating
this result gives $m(x)\geq c_\mu\mu(x)\,\forall
x,\mu\in \M_{comp}^{msr}$, i.e.\
$m\geqm\M_{comp}^{msr}$. Exponentiation of $\neg(o_3)$
implies
$m(x)\not\geqm\MM(x)\in\M_{enum}^{semi}$, i.e.\
$m\not\geqm\M_{enum}^{semi}$.
\qed

\subsection{Monotonicity of $m=2^{-\Km}$}

Monotonicity of $\Km$ is obvious from the definition of
$\Km$ and is the origin of calling $\Km$ monotone complexity:

\addtocounter{theorem}{-1}
\ftheorem{thmii}{ii) (Monotonicity of $m=2^{-\Km}$}{\hfill
\begin{list}{}{\parsep=1ex\itemsep=0ex\leftmargin=5ex\labelwidth=5ex}
\item[] $\Km(xy)\geq \Km(x)\in\SetN_0$,
$\quad 0<m(xy)\leq m(x)\in 2^{-\SetN_0}\leq 1=m(\epstr)$.
\end{list}
}

\subsection{Non-Semimeasure Property of $m=2^{-\Km}$}

While $m$ is monotone, it is not a semimeasure. The following
theorem shows and quantifies how the crucial semimeasure property
is violated for $m$ in an essential way.

\addtocounter{theorem}{-1}
\ftheorem{thmiii}{iii) (Non-Semimeasure property of $m=2^{-\Km}$}{\hfill
\begin{list}{}{\parsep=1ex\itemsep=0ex\leftmargin=5ex\labelwidth=5ex}
\item[$\scriptstyle\neg(1)$] If $x_{1:\infty}$ is computable,
then $\sum_{x_t}m(x_{1:t})\not\leq m(x_{<t})$ for almost all $t$,
\item[$\scriptstyle\neg(2)$] If $\;\Km(x_{1:t})=o(t)$, $\;$
then $\sum_{x_t}m(x_{1:t})\not\leq m(x_{<t})$ for most $t$.
\end{list}
}

\paradot{Remark} On the other hand, at least for computable environments,
multiplying Theorem~\ref{thmvi}$(vi_{1\&3})$ by $m(x_{<t})$ shows
that asymptotically the violation gets small, i.e.\
$\sum_{x_t}m(x_{1:t})\toinfty{t} m(x_{<t})$ for computable
$x_{1:\infty}$.

\paradot{Proof}
Simple violation of the semimeasure property can be inferred
indirectly from $m$ possessing properties $(i),(iv),\neg(vi)$ (see
Definition \ref{defbProp}) and Theorem \ref{thPredRel}$b$. To
prove $\bf\neg(iii_1)$ we first note that $\Km(x)<\infty$ for all
finite strings $x\in\X^*$, which implies $m(x_{1:n})>0$. Hence,
whenever $\Km(x_{1:n})=\Km(x_{<n})$, we have
$\sum_{x_n}m(x_{1:n})> m(x_{1:n})=m(x_{<n})$, a violation of the
semimeasure property. $\bf\neg(iii_2)$ now follows from
\bqan\ifjcss\nq\fi
  \#\{t\leq n: \sum_{x_t}m(x_{1:t})\leq m(x_{<t})\}
  &\;\leq\;& \#\{t\leq n: \Km(x_{1:t})\neq \Km(x_{<t})\}
\\
  &\;\leq\;& \sum_{t=1}^n [\Km(x_{1:t})-\Km(x_{<t})]
  \;=\; \Km(x_{1:n}),
\eqan
where we exploited $(ii)$ in the last inequality.
\qed

\subsection{Chain Rule for $m=2^{-\Km}$}

\addtocounter{theorem}{-1}
\ftheorem{thmiv}{iv) (Chain rule for $m=2^{-\Km}$}{\hfill
\begin{list}{}{\parsep=1ex\itemsep=0ex\leftmargin=5ex\labelwidth=5ex}
\item[$\scriptstyle(1)$]
$0 < m(x|y):= {m(yx)\over m(y)} \leq 1$.
\item[$\scriptstyle\neg(2)$]
If $m_|(x|y):= 2^{-\min_p\{\l(p):U(p,y)=x*\} }$, then
$\exists x,y : m(yx)\neq m_|(x|y)\cdot m(y)$.
\item[$\scriptstyle\neg(3)$]
$\Km(yx)=\Km_|(x|y)+\Km(y) \pm O(\log\,\l(xy))$.
\end{list}
}

\paradot{Remarks}
Line 1 shows that \mrcp\ can be satisfied by definition. With such a
definition, $m(x|y)$ is strictly positive like $M(x|y)$, but not
necessarily strictly less than $1$, unlike $M(x|y)$. Nevertheless
it is bounded by $1$ due to monotonicity of $m$, unlike for $k$
(see Theorem~\ref{thkProp}).
If a conditional monotone complexity $\Km_|=-\lb\,m_|$ is defined
similarly to the conditional Kolmogorov complexity $K_|$, then
\mrcp\ is only valid within logarithmic accuracy (lines 2 and 3).

\paranodot{Proof $\bf(iv_1)$} is immediate from $(ii)$.
$\bf\neg(iv_2)$ follows from the fact that equality does not even
hold within an additive constant, i.e.\ $\Km(yx)\not\equa
\Km(x|y)+\Km(y)$. The proof of the latter is similar to the one
for $K$ (see \cite{Li:97}).
$\bf\neg(iv_3)$ follows within $\log$ from $\Km=K+O(\log)$ and
Theorem~\ref{thkProp}$(iv)$.
\qed

\subsection{Enumerability of $m=2^{-\Km}$}

$m$ shares the obvious enumerability property with $M$ and
$\Km$ shares the obvious co-enumerability property with $K$:

\addtocounter{theorem}{-1}
\ftheorem{thmv}{v) (Enumerability of $m=2^{-\Km}$}{\hfill
\begin{list}{}{\parsep=1ex\itemsep=0ex\leftmargin=5ex\labelwidth=5ex}
\item[$\scriptstyle(1)$] $m$ is enumerable, i.e.\ lower semicomputable.
\item[$\scriptstyle(2)$] $\Km$ is co-enumerable, i.e.\ upper semicomputable.
\end{list}
}

\subsection{Convergence of $m=2^{-\Km}$}

\addtocounter{theorem}{-1}
\ftheorem{thmvi}{vi) (Convergence of $m=2^{-\Km}$}{\hfill
\begin{list}{}{\parsep=1ex\itemsep=0ex\leftmargin=5ex\labelwidth=5ex}
\item[$\scriptstyle(1)$] $\sum_{t=1}^n|1-m(x_t|x_{<t})|\leq \odt
\Km(x_{1:n})$, $\quad
m(x_t|x_{<t})\ifjcss\to\else\stackrel{fast}\longrightarrow\fi 1$ for comp.\
$x_{1:\infty}$.
\item[$\scriptstyle(2)$]
Indeed, $m(x_t|x_{<t})\neq 1$ at most $\Km(x_{1:\infty})$ times.
\item[$\scriptstyle(3)$]
$\sum_{t=1}^n\sum_{\bar x_t\neq x_t}
m(\bar x_t|x_{<t}) \leq 2^{\Km(x_{1:n})}$, $\qquad
m(\bar x_t|x_{<t})\ifjcss\to\else\stackrel{slow?}\longrightarrow\fi 0$ for
comp.\ $x_{1:\infty}$.
\item[$\scriptstyle(4)$]
$\sum_{t=1}^n\sum_{\bar x_t\neq x_t}
m(\bar x_t|x_{<t}) \leqm [\Km(x_{1:n})]^3$, $\quad
m(\bar x_t|x_{<t})\ifjcss\to\else\stackrel{fast?}\longrightarrow\fi 0$ for
comp.\ $x_{1:\infty}$.
\item[$\scriptstyle\neg(5)$]
$\forall s\, \exists\, U,x_{1:\infty} \,:\, \Km(x_{1:\infty})=s$ and
$\sum_{t=1}^\infty\sum_{\bar x_t\neq x_t}
m(\bar x_t|x_{<t}) \geq 2^s-2$.
\item[$\scriptstyle\neg(6)$]
$\exists\mu\in\M_{comp}^{msr}\setminus\M_{det} \,:\,
m_{(norm)}(x_t|x_{<t}) \;\;\not\!\!\!\toinfty{t} \mu(x_t|x_{<t})\,\forall x_{1:\infty}$
\end{list}
}

\paradot{Remarks}
Line 1 shows that the on-sequence predictive
properties of $m$ for deterministic computable environments are
excellent. The predicted $m$-probability\footnote{We say
``probability'' just for convenience, not forgetting that
$m(\cdot|x_{<t})$ is not a proper (semi)probability distribution.}
of $x_t$ given $x_{<t}$ converges rapidly to 1 for reasonably
simple $x_{1:\infty}$. A similar result holds for $M$.

The stronger result (second line), that $m(x_t|x_{<t})$ deviates
from 1 at most $\Km(x_{1:\infty})$ times, does not hold for $M$.

Note that without constraint on the predictive function $b$,
perfect on-sequence prediction could trivially be achieved by
defining $b_.(x'_t|x_{<t})\equiv 1$
$\forall x'_t$, which correctly predicts $x_t$ with ``probability'' 1.
But since we do not know the true outcome $x_t$ in advance, we
need to predict the probability of $x'_t$ well for all
$x'_t\in\X$. $m(|)$ also converges off-sequence for $\bar x_t\neq
x_t$ (to zero as it should be), but the bound (third line) is much
weaker than the on-sequence bound (first line), so rapid
convergence cannot be concluded, unlike for $M$, where
$M(x_t|x_{<t})\stackrel{fast}\longrightarrow 1$ implies $M(\bar
x_t|x_{<t})\stackrel{fast}\longrightarrow 0$, since
$\sum_{x'_t}M(x'_t|x_{<t})\leq 1$. Consider an
environment $x_{1:\infty}$ describable in 500 bits, then bound
$(vi_3)$ does not exclude $m(\bar x_t|x_{<t})$ from being 1
(maximally wrong) for all $t=1..2^{500}$; with asymptotic
convergence being of pure academic interest.

Line 4 presents a bound polynomial in $\Km$, which is
theoretically better than the exponential bound of line 3, but
there is a pitfall due to the hidden multiplicative constant.

Line 5 shows that for particular universal Turing machines this
constant can be exponentially large. Note that this does
not contradict the polynomial bound, since the multiplicative
constant $2^{c_U}$ is allowed to depend on $U$. For a reasonable
Turing machine, the compiler constant $c_U$ is of reasonable size,
but $2^{c_U}$ is unreasonably large.
Let $U'$ be a Turing machine which you regard
as reasonable. Then, for e.g.\ $s=64=O(1)$, the $U$ constructed in the
proof is as reasonable as $U'$ in the sense that a program of $U'$
needs only to be prefixed by a short 64 bit word to run on $U$
(the compiler constant between $U$ and $U'$ is small). In this
sense, there are {\em reasonable} Turing machines $U$ for which
$m$ makes the unreasonably large number of $2^{64}-2$ prediction
errors on the trivial sequence $0_{1:\infty}$, as we will show.

Line 6 shows that the situation is provably worse in the
probabilistic case. There are computable measures $\mu$ for which
neither $m(x_t|x_{<t})$ nor $m_{norm}(x_t|x_{<t})$ converge to
$\mu(x_t|x_{<t})$ for any $x_{1:\infty}$.
So while \cite[Thm.11]{Vitanyi:00} and \cite[Thm.5.2.3]{Li:97}
stating that $\mu(x_{t:t+l}|x_{<t}) \eqm m(x_{t:t+l}|x_{<t})$ for
$\mu$-random $x_{1:\infty}$ and fixed $l$ is correct, the
conclusion \cite[Cor.2]{Vitanyi:00} and \cite[Cor.5.2.2]{Li:97}
that ($m$ is good for prediction in the sense that) maximizing
$\mu(\cdot|x_{<t})$ is asymptotically equivalent to maximizing
$m(\cdot|x_{<t})$, is wrong. For this to be true we would need
convergence without multiplicative fudge, and which also holds
off-sequence, i.e.\ $m_{(norm)}(x'_t|x_{<t})\to\mu(x'_t|x_{<t})$,
but which $\neg(vi_6)$ just shows to fail (even on-sequence).

\paranodot{Proof {\bf (vi$_{1\&2}$)}}
$
  \quad\#\{t\leq n:m(x_t|x_{<t})\neq 1\}
  \;\leq\; \sum_{t=1}^n 2|1-m(x_t|x_{<t})|
  \;\leq\;
$
\beqn\textstyle
  \;\leq\; -\sum_{t=1}^n \lb m(x_t|x_{<t})
  \;=\; -\lb m(x_{1:n})
  \;=\; \Km(x_{1:n}).
\eeqn
In the first inequality we used $m:=m(x_t|x_{<t})\in
2^{-\SetN_0}$, hence $1\leq 2|1-m|$ for $m\neq 1$. In the second
inequality we used $1-m\leq-\odt\lb\,m$, valid for
$m\in[0,\odt]\cup\{1\}\supset 2^{-\SetN_0}$. In the first equality
we used (the $\log$ of) \mrcp\ $n$ times. For computable
$x_{1:\infty}$ we have $\sum_{t=1}^\infty|1-m(x_t|x_{<t})| \leq
\odt\Km(x_{1:\infty}) < \infty$, which implies $m(x_t|x_{<t})\to
0$ (fast if $\Km(x_{1:\infty})$ is of reasonable size). This shows
the first two lines of $(vi)$.

{\bf (vi$_{\bf 3}$)} Fix a sequence $x_{1:\infty}$ and
define $\Q:=\{x_{<t}\bar x_t \,:\, t\in\SetN,\,\bar x_t\neq
x_t\}$. $\Q$ is a prefix-free set of finite strings. For any such
$\Q$ and any semimeasure $\rho$, one can show that $\sum_{x\in
\Q}\rho(x)\leq 1$.\footnote{This follows from $1\geq\rho(A\cup B)\geq
\rho(A)+\rho(B)$ if $A\cap B=\{\}$, $\Gamma_x\cap\Gamma_y=\{\}$ if
$x$ not prefix of $y$ and $y$ not prefix of $x$, where
$\Gamma_x:=\{\omega:\omega_{1:\l(x)}=x\}$, hence $\sum_{x\in
\Q}\rho(\Gamma_x)\leq\rho(\bigcup_{x\in \Q}\Gamma_x)\leq 1$, and
noting that $\rho(x)$ is actually an abbreviation for
$\rho(\Gamma_x)$.} Since $M$ is a semimeasure lower-bounded by $m$
we get
\beqn
  \sum_{t=1}^n\sum_{\bar x_t\neq x_t} m(x_{<t}\bar x_t)
  \;\leq\; \sum_{t=1}^\infty\sum_{\bar x_t\neq x_t} m(x_{<t}\bar x_t)
  \;=\; \sum_{x\in \Q} m(x)
  \;\leq\; \sum_{x\in \Q} M(x)
  \;\leq\; 1.
\eeqn
With this, and using monotonicity of $m$ we get
\ifjcss
\bqan
  \sum_{t=1}^n\sum_{\bar x_t\neq x_t} m(\bar x_t|x_{<t})
  &\;=\;& \sum_{t=1}^n\sum_{\bar x_t\neq x_t}{m(x_{<t}\bar x_t)\over m(x_{<t})}
\\
  &\;\leq\;& \sum_{t=1}^n\sum_{\bar x_t\neq x_t}{m(x_{<t}\bar x_t)\over m(x_{1:n})}
  \;\leq\; {1\over m(x_{1:n})}
  \;=\; 2^{\Km(x_{1:n})}
\eqan
\else
\beqn
  \sum_{t=1}^n\sum_{\bar x_t\neq x_t} m(\bar x_t|x_{<t})
  = \sum_{t=1}^n\sum_{\bar x_t\neq x_t}{m(x_{<t}\bar x_t)\over
  m(x_{<t})}
  \leq \sum_{t=1}^n\sum_{\bar x_t\neq x_t}{m(x_{<t}\bar x_t)\over
  m(x_{1:n})}
  \leq {1\over m(x_{1:n})}
  = 2^{\Km(x_{1:n})}
\eeqn
\fi
Finally, for an infinite sum to be finite, its elements must
converge to zero.

{\bf (vi$_{\bf 4}$)} For $t\leq n$ we can bound
\beqn
  m(\bar x_t|x_{<t})
  \;\equiv\; {m(x_{<t}\bar x_t)\over m(x_{<t})}
  \;\leqm\; \Km^2(x_{<t}){M(x_{<t}\bar x_t)\over M(x_{<t})}
  \;\leq\; \Km^2(x_{1:n}) M(\bar x_t|x_{<t})
\eeqn
In the first inequality we exploited Theorem~\ref{thmo}$(o_2)$ in
the exponentiated form $M(x)/\Km^2(x)\leqm m(x)\leq M(x)$. In the
last inequality we used monotonicity of $m$. Using Theorem
\ref{thCUNSM} with $a=c=1$ and $b=M$ and $\KM\leq\Km$ we get
\beqn
  \sum_{t=1}^n\sum_{\bar x_t\neq x_t} m(\bar x_t|x_{<t})
  \;\leqm\; \Km^2(x_{1:n})\sum_{t=1}^n\sum_{\bar x_t\neq x_t} M(\bar x_t|x_{<t})
  \;\leq\; \ln 2 \!\cdot\! \Km^3(x_{1:n}).
\eeqn
Note that using $(o_1)$ instead of $(o_2)$ leads to a bound
$2^{c_\omega}\ln 2\cdot\Km(\omega)$, which for computable $\omega$
is also finite, but of unspecified magnitude due to the factor
$2^{c_\omega}$.

{\bf $\neg$(vi$_{\bf 5}$)}
Fix $s\in\SetN$ and let $t\in T:=\{1,...,2^s-2\}$. We define a
universal monotone Turing machine $U$ by $U(0^s)=0^\infty$ and
$U(q)=0^{t-1}1*$ for $q\in\B^s\setminus\{0^s,1^s\}$, where $t\in
T$ is the natural number represented by the $s$-bit string $q$
(any coding will do). Only for the purpose of making $U$
universal, we define $U(1^s p)=U'(p)$ for $p\in\B^*$ and $U'$
being some (other, e.g.\ your favorite) universal Turing machine.
Obviously the length of the shortest programs on $U$ for
$0_{1:\infty}$, $0_{<t}1$ and $0_{<t}$ is $s$, i.e.\
$\Km(0_{1:\infty})=\Km(0_{<t})=\Km(0_{<t}1)=s$, which implies
$m(1|0_{<t})=1$. So for $x_{1:\infty}=0_{1:\infty}$, we have
\beqn
  \sum_{t=1}^\infty\sum_{\bar x_t\neq x_t} m(\bar x_t|x_{<t})
  \;\geq\; \sum_{t=1}^{2^s-2} m(1|0_{<t})
  \;=\; 2^s-2,
\eeqn
which proves $\neg(iv_5)$.
Note that $m_{norm}(1|0_{<t})\geq{1\over|\X|}$,
i.e.\ save a factor of $|\X|$ the same lower bound holds for
$m_{norm}$. Note also that on-sequence prediction is perfect,
since $m(0|0_{<t})=1$ $\forall\,t\in\SetN$.

{\it Remark.} It is instructive to see why $M(\bar x_t|x_{<t})$
converges fast to 0 for this $U$: The single program of size $s$
for $0_{<t}1$ is outweighed by the $2^s-t$ programs of size $s$
for $0_{<t}$. Ignoring the contributions from $U'$, we have
$M(1|0_{<t})\approx{1\cdot 2^{-s}\over (2^s-t)\cdot
2^{-s}}={1\over 2^s-t}$, hence $\sum_{t=1}^{2^s-2} M(1|0_{<t})
\approx s\cdot\ln 2$.

{\bf $\neg$(vi$_{\bf 6}$)}
We show that the range of $m_{(norm)}$ is not dense in $[0,1]$ and
then choose a $\mu$ not in the closure of the range. For binary
alphabet $\X=\{0,1\}$, the proof is particularly simple:
We choose $\mu(1|x_{<t})={3\over 8}$, hence $\mu(0|x_{<t})={5\over
8}$. Since $m(x_t|x_{<t})\in 2^{-\SetN_0}=\{1,\odt,\odf,{1\over
8},...\}$, we have $|m(x_t|x_{<t})-\mu(x_t|x_{<t})|\geq{1\over
8}\,\forall t,\,\forall x_{1:\infty}$. Similarly for
\beqn
  m_{norm}(x_t|x_{<t}) \;=\;
  {m(x_t|x_{<t})\over m(0|x_{<t})+m(1|x_{<t})}
  \;\in\; \left\{ {2^{-n}\over 2^{-n}+2^{-m}}:n,m\!\in\!\SetN_0 \right\} \;=
\eeqn\vspace{-2ex}
\beqn
    =\; \left\{ {1\over 1+2^z}:z\!\in\!\SetZ\right\}
  \;=\; {1\over 1+2^\SetZ}
  \;=\; \left\{ ...,{1\over 9},{1\over 5},{1\over 3},{1\over 2},
       {2\over 3},{4\over 5},{8\over 9},... \right\}
\eeqn
we choose $\mu(1|x_{<t})=1-\mu(0|x_{<t})={5\over 12}$, which
implies $|m_{norm}(x_t|x_{<t})-\mu(x_t|x_{<t})|\geq{1\over
12}$ $\forall t$, $\forall x_{1:\infty}$.

Consider now a general alphabet $\X=\{1,...,|\X|\}$, and the
unnormalized $m$ first.
If $|\X|$ is not a power of 2 we define
$\mu(x_t|x_{<t})=|\X|^{-1}$. If $|\X|$ is a power of 2 we define
$\mu(x_t|x_{<t})={4\over 3}|\X|^{-1}$ for even $x_t$ and
$\mu(x_t|x_{<t})={2\over 3}|\X|^{-1}$ for odd $x_t$. $\mu$ is a
measure, $0\neq\mu(x_t|x_{<t})\not\in 2^{-\SetN_0}$, but
$m(x_t|x_{<t})\in 2^{-\SetN_0}$. The only cluster\footnote{A
point $p\in\SetR^n$ is called a cluster point of a set
$\S\subseteq\SetR^n$, if every open set of $\SetR^n$ which
contains $p$, intersects $\cal S$.} point of $2^{-\SetN_0}$ is
$0$, since $0\neq\mu\not\in 2^{-\SetN_0}$ there exists
$\gamma>0$ such that $(\mu-\gamma,\mu+\gamma)\cap 2^{-\SetN_0}=\{\}$,
hence $|m(x_t|x_{<t})-\mu(x_t|x_{<t})|\geq\gamma\,\forall
t,\,\forall x_{1:\infty}$ for some $\gamma>0$.

For $m_{norm}$ we proceed as follows: With
$z_i:=\Km(1|x_{<t})-\Km(i|x_{<t})\in\SetZ$, we have
$m_{norm}(1|x_{<t})^{-1}=1+\sum_{i=2}^{|\X|}2^{z_i}$. We define
$\S :=\{1+m_2+...+m_{|\X|} \,:\, m_i\in 2^\SetZ\cup\{0\}\forall
i\}\not\ni 0$ and $\I :=\S ^{-1}=\{x^{-1} \,:\, x\in\S \}$. By
construction, $m_{norm}(1|x_{<t})\in\cal I$, and by symmetry also
$m_{norm}(x_t|x_{<t})\in\cal I$. The cross product
$\I^{|\X|}:=\I\times\stackrel{|\X| times}{...........}\times\I$ is
a closed and countable set, since $2^\SetZ\cup\{0\}$ is closed
and countable, and finite sums, inversions, and cross products of
closed/countable sets, are closed/countable.$\!$\footnote{W.r.t.\
standard topology on $\SetR^n$.} With $\Delta:=\{\v v\in\SetR^{|\X|} \,:\,0<v_i<1,\;\sum_{i=1}^{|\X|} v_i=1\}$ being the open
$|\X|-1$ dimensional simplex, we have
$m_{norm}(\cdot|x_{<t})\in\I^{|\X|}\cap\Delta$ (e.g.\
$\I^2\cap\Delta=\{({1\over 1+2^z},{1\over 1+2^{-z}}) : z\in\SetZ\}$). Since $\Delta\setminus\I^{|\X|}$ is open and nonempty (due
to countability of $\I^{|\X|}$), there exists
$\mu(\cdot|x_{<t})\in\Delta\setminus\I^{|\X|}$ and a Box$:=\{\v
v:|v_i-\mu(i|x_{<t})|<\gamma\}$ of sufficiently small size
$\gamma>0$ surrounding $\mu$, such that Box$\cap\I^{|\X|}=\{\}$,
which implies the desired result
$|m(x_t|x_{<t})-\mu(x_t|x_{<t})|\geq\gamma$.

{\it Remark.} There is an easy proof for the weaker statement
$m_{norm}(x'_t|x_{<t})\not\to\mu(x'_t|x_{<t})$, where $x'_t$ may
be off-sequence: For $\mu(0|x_{<t})={1\over 4}=1-\mu(1|x_{<t})$ we
have ${\mu(1|x_{<t})\over\mu(0|x_{<t})}=3\not\in 2^\SetZ$,
while ${m_{norm}(1|x_{<t})\over m_{norm}(0|x_{<t})}\in 2^\SetZ$. This implies that the posterior of $m_{norm}$ cannot be too
close to the posterior of $\mu$ for {\em all} $x'_t$, i.e.\ $\exists x'_t$ and $c>0$ :
$|m_{norm}(x'_t|x_{<t})-\mu(x'_t|x_{<t})|\geq c$ ($c={1\over 20}$
possible). One advantage of this proof is that it also goes
through for infinite alphabet $\X$.
\qed

\subsection{Self-optimization of $m=2^{-\Km}$}

\addtocounter{theorem}{-1}
\ftheorem{thmDProp}{vii) (Self-optimization of $m=2^{-\Km}$}{\hfill
\begin{list}{}{\parsep=1ex\itemsep=0ex\leftmargin=5ex\labelwidth=5ex}
\item[$\scriptstyle(1)$]
$l_t^{\smash{\Lambda_m}}(x_{<t}) \ifjcss\to\else\;\stackrel{slow?}\longrightarrow\;\fi
l_t^{\smash{\Lambda_\omega}}:=\arg\min_{y_t}\ell_{x_ty_t}$
if $\omega\equiv x_{1:\infty}$ is computable.
\item[$\scriptstyle(2)$]
$\Lambda_m=\Lambda_{m_{norm}}$, i.e.\
$y_t^{\smash{\Lambda_m}}=y_t^{\smash{\Lambda_{m_{norm}}}}$ and
$l_t^{\smash{\Lambda_m}}=l_t^{\smash{\Lambda_{m_{norm}}}}$.
\item[$\scriptstyle\neg(3)$]
$\forall |\Y|>2\,\exists\ell,
\mu \;:\; {l_t^{\smash{\Lambda_m}}/l_t^{\smash{\Lambda_\mu}}}
= c > 1\,\forall t\quad$
$(c={6\over 5}-\eps$ possible$)$.
\item[$\scriptstyle\neg(4)$]
$\exists\ell,\mu \;:\; {l_t^{\smash{\Lambda_m}}/l_t^{\smash{\Lambda_\mu}}}
= c > 1$ for many $t$
with $\mu$-probability $\geq\odt$  \ifjcss \\ \fi
$(c=\sqrt{2}-\eps$ possible$)$.
\item[$\scriptstyle\neg(5)$]
$\forall$ non-degenerate\footnote{A formal definition of {\em non-degenerate} is
given in the remarks after the theorem.} $\ell\;\exists U,\mu \;:\;
{l_t^{\smash{\Lambda_m}}/l_t^{\smash{\Lambda_\mu}}} \;\;\not\!\!\!\toinfty{t} 1$
with high probability.
\end{list}
}

\paradot{Remarks} Since $(vi)$ implies $(vii_1)$ by continuity,
we have convergence of the instantaneous losses for computable
environments $x_{1:\infty}$, but since convergence off-sequence is
potentially slow, the convergence of the losses to optimum is
potentially slow.

Non-convergence $\neg(vi_6)$ in probabilistic environments does
not necessarily imply that $\Lambda_m$ is not self-optimizing,
since different predictive functions can lead to the same
predictor $\Lambda$. But $\neg(vii_4)$ shows that $\Lambda_m$ is
not self-optimizing even in Bernoulli environments $\mu$ for
particular losses $\ell$ with probability $\geq\odt$.

Interestingly, excluding binary action alphabets allows for a
stronger for-sure statement $\neg(vii_3)$.

In $\neg(vii_5)$, non-self-optimization is shown for {\em any}
{\em non-degenerate loss function} (especially  for the error
loss, cf.\ (\ref{eqMDLk})), for specific choices of the universal
Turing machine $U$. Loss $\ell$ is defined to be non-degenerate
{\em iff} $\bigcap_{x\in\X}\{\tilde y\,:\,\ell_{x\tilde
y}=\min_y\ell_{xy}\} = \{\}$. Assume the contrary that a {\it
single} action $\tilde y$ is optimal for {\it every} outcome $x$,
i.e.\ that ($\arg\min_y$ can be chosen such that)
$\arg\min_y\ell_{xy}=\tilde y\,\forall x$. This implies
$y_t^{\smash{\Lambda_\rho}}=\tilde y\,\forall\rho$, which implies
$l_t^{\smash{\Lambda_m}}/l_t^{\smash{\Lambda_\mu}}\equiv 1$. So
the non-degeneracy assumption is necessary (and sufficient).

\paranodot{Proof $\bf(vii_1)$}
follows from $(vi_{1\&3})$ and Theorem~\ref{thPredRel}d.

$\bf(vii_2)$
That normalization does not affect the predictor,
follows from the definition of $y_t^{\smash{\Lambda_\rho}}$ (\ref{xlrdef})
and the fact that $\arg\min()$ is not affected by scaling its
argument.

$\bf\neg(vii_3)$
Non-convergence of $m$ does not necessarily imply non-convergence
of the losses. For instance, for $\X=\Y=\{0,1\}$, and
$\omega'_t:=1/0$ for
$\mu(1|x_{<t}){>\atop<}\gamma := {\ell_{01}-\ell_{00}\over
\ell_{01}-\ell_{00}+\ell_{10}-\ell_{11}}$, one can show
that $y_t^{\smash{\Lambda_\mu}}=y_t^{\smash{\Lambda_{\omega'}}}$, hence
convergence of $m(x_t|x_{<t})$ to 0/1 and not to $\mu(x_t|x_{<t})$
could nevertheless lead to correct predictions.

Consider now $x\in\X=\{0,1\}$,
$y\in\Y=\{0,1,2\}$. To prove $\neg(vii_3)$ we define a loss function such
that $y_t^{\smash{\Lambda_\mu}}\neq y_t^{\smash{\Lambda_\rho}}$ for any $\rho$
with same range as $m_{norm}$ and for some $\mu$. The loss
function $\ell_{x0}=x$, $\ell_{x1}={3\over 8}$, $\ell_{x2}={2\over
3}(1-x)$, and $\mu:=\mu(1|x_{<t})={2\over 5}$ will do. The
$\rho$-expected loss under action $y$ is $l_\rho^y:=\sum_{x_t=0}^1
\rho(x_t|x_{<t})\ell_{x_t y}$; $l_\rho^0=\rho$, $l_\rho^1={3\over
8}$, $l_\rho^2={2\over 3}(1-\rho)$ with $\rho:=\rho(1|x_{<t})$
(see Figure \ref{figLoss}).
\def\spfrac#1#2{{{^{#1}\!/\!_{#2}}}}
\def\figlossdef{
\begin{picture}(200,150)(-50,-10)
\thicklines
\put(0,0){\vector(1,0){140}}\put(130,-2){\makebox(0,0)[lt]{$\rho$}}
\put(-1,120){\line(1,0){2}}\put(-2,120){\makebox(0,0)[rc]{$1$}}
\put(0,0){\vector(0,1){135}}\put(-2,125){\makebox(0,0)[rb]{$l_\rho^y$}}
\put(120,-1){\line(0,1){2}}\put(120,-2){\makebox(0,0)[ct]{$1$}}
\put(0,0){\dashbox(120,120){}}
\put(1.85,0){\circle*{1.3}}
\put(3.64,0){\circle*{1.5}}
\put(7.06,0){\circle*{1.8}}
\put(13.33,0){\circle*{2}}
\put(24,0){\circle*{2}}\put(24,-2){\makebox(0,0)[ct]{$\spfrac{1}{5}$}}
\put(40,0){\circle*{2}}\put(40,-2){\makebox(0,0)[ct]{$\spfrac{1}{3}$}}
\put(60,0){\circle*{2}}\put(60,-2){\makebox(0,0)[ct]{$\spfrac{1}{2}$}}
\put(80,0){\circle*{2}}\put(80,-2){\makebox(0,0)[ct]{$\spfrac{2}{3}$}}
\put(96,0){\circle*{2}}\put(96,-2){\makebox(0,0)[ct]{$\spfrac{4}{5}$}}
\put(106.67,0){\circle*{2}}
\put(112.94,0){\circle*{1.8}}
\put(116.36,0){\circle*{1.5}}
\put(118.15,0){\circle*{1.3}}
\put(0,0){\line(1,1){45}}
\put(45,45){\line(1,0){7.5}}
\put(120,0){\line(-3,2){67.5}}
\thinlines
\put(0,0){\line(1,1){120}}
\put(-2,0){\makebox(0,0)[rc]{$\ell_{00}=0$}}
\put(119,120){\line(1,0){2}}\put(122,120){\makebox(0,0)[lc]{$1=\ell_{10}$}}
\put(100,100){\makebox(0,0)[lt]{$l_\rho^0$}}
\put(0,45){\line(1,0){120}}
\put(-1,45){\line(1,0){2}}\put(-2,45){\makebox(0,0)[rc]{$\ell_{01}=\,^3\!\!/\!_8$}}
\put(119,45){\line(1,0){2}}\put(122,45){\makebox(0,0)[lc]{$^3\!\!/\!_8=\ell_{11}$}}
\put(100,47){\makebox(0,0)[lb]{$l_\rho^1$}}
\put(120,0){\line(-3,2){120}}
\put(-1,80){\line(1,0){2}}\put(-2,80){\makebox(0,0)[rc]{$\ell_{02}=\,^2\!\!/\!_3$}}
\put(119,0){\line(1,0){2}}\put(122,2){\makebox(0,0)[lb]{$0=\ell_{12}$}}
\put(100,13){\makebox(0,0)[lb]{$l_\rho^2$}}
\put(40,0){\dashbox(20,120){}}
\put(0,40){\dashbox(60,0){}}
\put(-1,40){\line(1,0){2}}
\put(-2,39){\makebox(0,0)[rt]{$^1\!\!/\!_3$}}
\put(0,0){\dashbox(48,48){}}
\put(-1,48){\line(1,0){2}}
\put(-2,50){\makebox(0,0)[rb]{$^2\!\!/\!_5$}}
\put(48,-1){\line(0,1){2}}
\put(48,-2){\makebox(0,0)[ct]{$\spfrac{2}{5}$}}
\end{picture}
}
\begin{figure}[htb]
\ffigurex{figLoss}{Example loss used in proof of Theorem
\ref{thmDProp}$\neg(vii)$}{The $\rho$-expected expected losses
$l_\rho^y$ under actions $y\in\Y=\{0,1,2\}$ for $\X=\{0,1\}$ and
loss function $\ell_{00}=\ell_{12}=00$,
$\ell_{01}=\ell_{11}={3\over 8}$, $\ell_{02}={2\over 3}$,
and $\ell_{10}=1$ are displayed as solid lines.
}{%
\small\unitlength=0.5mm
\begin{center}
\thicklines\figlossdef
\end{center}}
\end{figure}
Since $l_\mu^0=l_\mu^2={2\over 5}>{3\over 8}=l_\mu^1$, we have
$y_t^{\smash{\Lambda_\mu}}=1$ and $l_t^{\smash{\Lambda_\mu}}=l_\mu^1={3\over 8}$.
For $\rho\leq{1\over 3}$, we have
$l_\rho^0<l_\rho^1<l_\rho^2$, hence $y_t^{\smash{\Lambda_\rho}}=0$ and
$l_t^{\smash{\Lambda_\rho}}=l_\mu^0={2\over 5}$.
For $\rho\geq\odt$, we have
$l_\rho^2<l_\rho^1<l_\rho^0$, hence $y_t^{\smash{\Lambda_\rho}}=2$ and
$l_t^{\smash{\Lambda_\rho}}=l_\mu^2={2\over 5}$.
Since $m_{norm}\not\in({1\over 3},\odt)$, $\Lambda_{m_{norm}}$
predicts $0$ or $2$, hence
$l_t^{\smash{\Lambda_m}}=l_\mu^{0/2}={2\over 5}$. Since
$\Lambda_{m_{norm}}=\Lambda_m$, this shows that
$l_t^{\smash{\Lambda_m}}/l_t^{\smash{\Lambda_\mu}}={16\over
15}>1$. The constant ${16\over 15}$ can be enlarged to ${6\over
5}-\eps$ by setting $\ell_{x1}={1\over 3}+\eps$ instead of
${3\over 8}$.

For $\Y=\{0,...,|\Y|-1\}$, $|\Y|>3$, we extend the loss function
by defining $\ell_{xy}=1$ $\forall y\geq 3$, ensuring that actions
$y\geq 2$ are never favored.
For $\X=\{0,...,|\X|-1\}$, $|\X|>2$,
we extend $\mu$ and define $\mu(x_t|x_{<t})=0$ $\forall x_t\geq
2$. Furthermore, we define $\ell_{xy}=0$ for $x\geq 2$ and $y<3$.
This ensures that the extra components of $m_{norm}(x_t|x_{<t})$
with $x_t\geq 2$ do not contribute to $l_{m_{norm}}^y$. Finally,
and this is important, we define, solely for the purpose of this
proof, $m_{norm}(x_t|x_{<t})={m(x_t|x_{<t})\over
m(0|x_{<t})+m(1|x_{<t})}$, such that
$m_{norm}(0|x_{<t})+m_{norm}(1|x_{<t})=1$ (rather than
$\sum_{x_t=0}^{|\X|-1} m_{norm}(x_t|x_{<t})=1$) (Normalization
influences the analysis, but not the result). With these
extensions, the analysis of the $|\X|=2$, $|\Y|=3$ case applies,
which finally shows $\neg(vii)$.
In general, a non-dense range of $\rho(x_t|x_{<t})$ implies
$l_t^{\smash{\Lambda_\rho}}\not\to l_t^{\smash{\Lambda_\mu}}$,
provided $|\Y|\geq 3$.

$\bf\neg(vii_4)$ We consider binary $\X=\Y=\B$ first. The proof
idea and notation is similar to $\neg(vii_3)$. We choose a
$\mu:=\mu(1|x_{<t})\not\in{1\over 1+2^\SetZ}$. Let
$a,b\in{1\over 1+2^\SetZ}$ with $a<\mu<b$ be the nearest (to
$\mu$) possible values of $m_{norm}\in{1\over 1+2^\SetZ}$. For
a fixed sequence $x_{1:\infty}$, we have either $m(1|x_{<t})\leq
a$ for (infinitely) many $t$ or $m(1|x_{<t})\geq b$ for
(infinitely) many $t$ (or both). Choosing $x_{1:\infty}$ at
random, we have either $m(1|x_{<t})\leq a$ for many $t$ with
$\mu$-probability $\geq\odt$ or $m(1|x_{<t})\geq b$ for many $t$
with $\mu$-probability $\geq\odt$ (or both).
Assume the former; for the latter the proof is analogous. We
consider a loss function such that $l_a^1>l_a^0$ and
$l_\mu^1<l_\mu^0$. Then also $l_m^1>l_m^0$ whenever $m\leq a$,
which is the case for many $t$ by assumption. Hence
$l_t^{\smash{\Lambda_m}}/l_t^{\smash{\Lambda_\mu}}=l_\mu^0/l_\mu^1=c>1$.
For instance, choose $\mu=\sqrt{2}-1$ and $\ell_{00}=0$ and
$\ell_{10}=1$ ($\Rightarrow l_\rho^0=\rho$). We get
$c=\sqrt{2}-O(\eps)$ by choosing $\ell_{01}=\odt+\eps$ and
$\ell_{11}=0$ ($\Rightarrow l_\rho^1=(\odt+\eps)(1-\rho)$) in the
former case with $a={1\over 3}$ (and $\ell_{01}=1-\eps$ and
$\ell_{11}=0$ ($\Rightarrow l_\rho^1=(1-\eps)(1-\rho)$) in the
latter case with $b=\odt$ and $l_b^1<l_b^0$ and $l_\mu^1>l_\mu^0$).
The generalization to general $\X$ and $\Y$ can be performed
similarly to $\neg(vii_3)$.

$\bf\neg(vii_5)$
We first present a simple proof for a particular loss function and
$\X=\Y=\B$, which contains the main idea also used to prove the
general result. We define a monotone Turing machine $U$ by
$U(1x0)=x0$ for all $x\in\X^*$. More precisely, if the first bit
of the input tape of $U$ contains 1, $U$ copies the half-infinite
input tape (without the first 1) to the output tape, but always
withholds the output until a $0$ appears. We have
$\Km(x1)=\Km(x10)=\l(x)+2=\Km(x0)+1$, which implies
$m_{norm}(1|x)={1\over 3}$ and $m_{norm}(0|x)={2\over 3}$. For the
loss function $\ell_{00}=\ell_{11}=0$, $\ell_{10}=1$,
$\ell_{01}={2\over 3}$ and a Bernoulli($\odt$) process
$\mu(x_t|x_{<t})=\odt$ we get $l_\mu^1=\odt\cdot{2\over
3}<\odt=l_\mu^0$ and $l_{m_{norm}}^1={2\over 3}\cdot{2\over
3}>{1\over 3}=l_{m_{norm}}^0$, hence
$l_t^{\smash{\Lambda_m}}/l_t^{\smash{\Lambda_\mu}}=l_\mu^0/l_\mu^1={3\over
2}>1$. $U$ is not yet universal. We make $U$ universal by
additionally defining $U(0^{s+1}p)=U'(p)$ for some (large, but
reasonable) $s\in\SetN$ and some (other) universal monotone TM
$U'$. We have to check whether this can alter (lower) the monotone
complexity. Fix $n$. Every $x$ of length $n$ has description $1x0$
of length $n+2$, so $U'$ only matters if $U'(p)=x*$ for some $p$
of length $<n-s+1$. Since there are at most $2^{n-s}$ minimal
programs of length $\leq n-s$, the fraction of problematic $x$ is
at most $2^{-s}$. Since $x$ is drawn at random, the loss ratio
$l_t^{\smash{\Lambda_m}}/l_t^{\smash{\Lambda_\mu}}={3\over 2}$,
hence, holds with high probability ($\geq 1-2^{-s}$). A martingale
argument (see below) shows that this implies
$l_t^{\smash{\Lambda_m}}/l_t^{\smash{\Lambda_\mu}}
\;\;\not\!\!\!\toinfty{t} 1$ (w.h.p.).

We now consider the case of general loss and alphabets.
In case where ambiguities in the choice of $y$ in
$\arg\min_y\ell_{xy}$ matter we consider the set of solutions
$\{\arg\min_y\ell_{xy}\}:=\{\tilde y:\ell_{x\tilde
y}=\min_y\ell_{xy}\}\neq\{\}$.
By assumption, $\ell$ is non-degenerate, i.e.\
$\bigcap_{x\in\X}\{\arg\min_y\ell_{xy}\}
= \{\}$. Let $\X_m$ be a minimal subset of $\X$ with
$\bigcap_{x\in\X_m}\{\arg\min_y\ell_{xy}\} = \{\}$. Take any
decomposition $\X_0\dot\cup\X_1=\X_m$ with $\X_0\neq\{\}\neq\X_1$,
which is possible, since $|\X_m|\geq 2$. We have
$\Y_i:=\bigcap_{x\in\X_i}\{\arg\min_y\ell_{xy}\} \neq \{\}$, since
$\X_m$ is minimal. Further, $\Y_0\cap\Y_1=\Y_m=\{\}$. It is
convenient to choose $|\X_1|=1$. W.l.g.\ we assume $\X_1=\{1\}$.

Define some $\Q\subset\B^s$, $|\Q|=|\X_0|$, a bijection
$b:\Q\to\X_0$, and a one-to-one (onto $\A$) decoding function
$d:\B^s\to \A$ with $\A=\X_0 1^s\cup 1\B^s\setminus
1\Q\subset\X^{s+1}$ as $d(x)=b(x)1^s$ for $x\in\Q$ and
$d(x)=1x$ for $x\in\B^s\setminus\Q$ with a large $s\in\SetN$ to
be determined later. We extend $d$ to $d:(\B^s)^*\to \A^*$ by
defining $d(z_1...z_k)=d(z_1)...d(z_k)$ for $z_i\in\B^s$ and
define the inverse coding function $c:\A\to\B^s$ and its extension
$c:\A^*\to(\B^s)^*$ by $c=d^{-1}$.

Roughly, $U$ is defined as
$U(1p_{1:sn}q)=d(p_{1:sn})b(q)1^s$ for $q\in\Q$. More precisely, if the
first bit of the binary input tape of $U$ contains 1, $U$ decodes
the successive blocks of size $s$, but always withholds the output
until a block $q\in\Q$ appears. $U$ is obviously monotone.
Universality will be guaranteed by defining $U(0p)$ appropriately,
but for the moment we set $U(0p)=\epstr$.
It is easy to see that for $x\in \A^*$ we have
\ifjcss
\beqn
\begin{array}{rcl}
\Km(x x_0) =& \Km(x x_0 1^s) &=\;\l(c(x))+s+1 \qmbox{for $x_0\in\X_0$,} \\
\Km(x1) =& \Km(x1z0_{1:s+1}) &=\;\l(c(x))+2s+1 \qmbox{for any $z\in\B^s\setminus\Q$,} \\
\Km(xy) =&                   &=\;\infty \qmbox{for any $y\in\X\setminus(\X_0\cup\{1\})$.}
\end{array}
\eeqn\vspace{-4ex}\beq\label{eqEUKm}\eeq
\else
\beq\label{eqEUKm}
\begin{array}{rcl}
\Km(x x_0) =& \Km(x x_0 1^s) &=\;\l(c(x))+s+1 \qmbox{for $x_0\in\X_0$,} \\
\Km(x1) =& \Km(x1z0_{1:s+1}) &=\;\l(c(x))+2s+1 \qmbox{for any $z\in\B^s\setminus\Q$,} \\
\Km(xy) =&                   &=\;\infty \qmbox{for any $y\in\X\setminus(\X_0\cup\{1\})$.}
\end{array}
\eeq
\fi
Hence, $m_{norm}(x_0|x)=[|\X_0|+2^{-s}]^{-1}\toinfty{s} 1$ and
$m_{norm}(1|x)=[2^s|\X_0|+1]^{-1}\toinfty{s} 0$ and
$m_{norm}(y|x)=0$. For $t-1\in(s+1)\SetN$ we get
$l_m^{y_t}:=\sum_{x_t}m_{norm}(x_t|x_{<t}) \ell_{x_t
y_t}\toinfty{s}{1\over|\X_0|}\sum_{x_t\in\X_0}\ell_{x_t y_t}$.
This implies
\beq\label{eqEUm}\ifjcss\nq\fi
  y_t^{\smash{\Lambda_m}}\in\{\arg\min_{y_t}l_m^{y_t}\}
  \subseteq \{\arg\min_y{\textstyle{1\over|\X_0|}}\!\!\sum_{x\in\X_0}\ell_{xy}\}
  = \bigcap_{x\in\X_0}\{\arg\min_y\ell_{xy}\} \equiv \Y_0.
\eeq
Inclusion $\subseteq$ holds for sufficiently large finite $s$.
Equality $=$ holds, since the set of points which are global
maxima of a linear average of functions coincides with the set of
points which simultaneously maximize all these
functions, if the latter is nonempty.

We now define $\mu(z)=|\A|^{-1}=2^{-s}$ for $z\in \A$ and $\mu(z)=0$
for $z\in\X^{s+1}\setminus \A$, extend it to $\mu(z_1...z_k) :=
\mu(z_1)\cdot...\cdot\mu(z_k)$ for $z_i\in\X^{s+1}$, and finally
extend it uniquely to a measure on $\X^*$ by
$\mu(x_{<t}):=\sum_{x_{t:n}}\mu(x_{1:n})$ for $\SetN\ni t\leq
n\in(s+1)\SetN$. For $x\in \A^*$ we have
$\mu(x_0|x)=\mu(x_0)=\mu(x_0 1^s)=2^{-s}\toinfty{s} 0$ and
$\mu(1|x)=\mu(1)=\sum_{y\in\X^s}\mu(1y)=\sum_{y\in
\B^s\setminus\Q}\mu(1y) =(2^s-|Q|)\cdot
2^{-s}=1-|X_0|2^{-s}\toinfty{s} 1$. For $t-1\in(s+1)\SetN$ we get
$l_\mu^{y_t}:=\sum_{x_t}\mu(x_t|x_{<t}) \ell_{x_t
y_t}\toinfty{s}\ell_{1 y_t}$. This implies
\beq\label{eqEUmu}\ifjcss\nq\nq\fi
  y_t^{\smash{\Lambda_\mu}}\in\{\arg\min_{y_t}l_\mu^{y_t}\}
  \subseteq \{\arg\min_y\ell_{1y}\} \equiv \Y_1
  \quad\mbox{for sufficiently large finite $s$}.
\eeq
Since $\Y_0\cap\Y_1=\{\}$,
(\ref{eqEUm}) and (\ref{eqEUmu}) imply
$y_t^{\smash{\Lambda_m}}\neq y_t^{\smash{\Lambda_\mu}}$, which implies
$l_t^{\smash{\Lambda_m}}\neq l_t^{\smash{\Lambda_\mu}}$ (otherwise the choice
$y_t^{\smash{\Lambda_m}}= y_t^{\smash{\Lambda_\mu}}$ would have been possible),
which implies
$l_t^{\smash{\Lambda_m}}/l_t^{\smash{\Lambda_\mu}}=c>1$ for $t-1\in(s+1)\SetN$, i.e.\ for (infinitely) many $t$.

What remains to do is to extend $U$ to a universal Turing machine.
We extend $U$ by defining $U(0zp)=U'(p)$ for any $z\in\B^{3s}$,
where $U'$ is some universal Turing machine.
Clearly, $U$ is now universal. We have to show that this extension
does not spoil the preceding consideration, i.e.\ that the shortest
code of $x$ has sufficiently often the form $1p$ and sufficiently
seldom the form $0p$. Above, $\mu$ has been chosen in such a way that
$c(x)$ is a Shannon-Fano code for $\mu$-distributed strings,
i.e.\ $c(x)$ is with high $\mu$-probability a shortest code of $x$.
More precisely, $\l(c(x))\leq\Km_T(x)+s$ with $\mu$-probability at
least $1-2^{-s}$, where $\Km_T$ is the monotone complexity w.r.t.\
any decoder $T$, especially $T=U'$. This implies
$
  \min_p\{\l(0p) : U(0p)=x*\}
  = 3s+1+\Km_{U'}(x)
  \geq 3s+1+\l(c(x))-s
  > \l(c(x))+s+1
  \geq \min_p\{\l(1p) : U(1p)=x*\},
$
where the first $\geq$ holds with high probability ($1-2^{-s}$)
and the last $\geq$ holds with $\mu$-probability 1. This shows
that the expressions (\ref{eqEUKm}) for $\Km$ are with high
probability (w.h.p.) not affected by the extension of $U$.
Altogether this
shows ${l_t^{\smash{\Lambda_m}}/l_t^{\smash{\Lambda_\mu}}}=c>1$ w.h.p.

A martingale argument can strengthen this result to yield
non-self\-opti\-mizing\-ness. For
$z_t:={M(\omega_{1:t})\over\mu(\omega_{1:t})}$ we have $z_0=1$,
$\E[z_t]\leq 1$, and $\E[z_t|\omega_{<t}]\leq z_{t-1}$, hence
$-z_t$ is a non-positive semi-martingale.
\cite[Thm.$4.1s$,p324]{Doob:53} now implies that
$z_\infty:=\lim_{t\to\infty} z_t$ exists w.p.1 and
$\E[z_\infty]\leq \lim_{t\to\infty}\E[z_t]\leq 1$. The Markov
inequality now yields
\beqn
  \P[\lim_{t\to\infty}(\KM(\omega_{1:t})+\lb\mu(\omega_{1:t}))\leq -s]=
  \P[z_\infty\geq 2^s]\leq
  2^{-s}\E[z_\infty]\leq 2^{-s}.
\eeqn
Substituting $\KM\leq\Km\leadsto\Km_{U'}$ and
$-\lb\mu(x)=\l(c(x))$ this shows that
$\l(c(\omega_{1:t}))\leq\Km_{U'}(\omega_{1:t})+s$ for almost all
$t\in(s+1)\SetN$ with probability $\geq 1-2^{-s}$.
Altogether this shows ${l_t^{\smash{\Lambda_m}}/l_t^{\smash{\Lambda_\mu}}}
\;\;\not\!\!\!\toinfty{t} 1$ w.h.p.
\qed

\section{Further Results}\label{secFurther}
\label{secPPK}\label{secAlEU}\label{secSMDL}

\paradot{Predictive Properties of $k=2^{-K}$}
We briefly discuss the predictive properties of the prefix
Kolmogorov complexity $K$. We will be very brief, since $K$
completely fails for predictive purposes, although $K$ is close to
$\KM$ within an additive logarithmic term.

\ftheorem{thkProp}{Properties of $k=2^{-K}$}{
For $b=k=2^{-K}$, where $K$ is the prefix Kolmogorov complexity,
the following properties of Definition~\ref{defbProp} are
satisfied/violated: $(o)$ $\KM(x)\leq K(x)\leq \KM(x)+2\lb K(x)$.
$(i),(ii),(iii)$ are violated. $(iv)$ is satisfied only for $k.=k$
$\;$ For $k.=k_|$ $(iv)$ is only satisfied to logarithmic order. In
any case $(vi)$ and $(vii)$ can be violated for deterministic as
well as probabilistic $\mu\in\M_{comp}^{msr}$. $(v)$ is satisfied.
}

\paradot{Proof sketch}
$(o)$ Similar to proof of Theorem 3.4 in \cite{Zvonkin:70}.
$\neg(i)$ for deterministic
$\mu\in\M_{comp}^{msr}$ with $\mu(0_{1:n})=1$, we have
$k(0_{1:n})\to 0\not\geqm  1=\mu(0_{1:n})$,
since $K(\omega_{1:n})\toinfty{n}\infty$ $\forall\omega$.
$\neg(ii)$, since $K(0_{1:n})\equa K(n)\geq\lb\,n$ for most $n$,
but $\leqa  2\lb\log\,n$ for $n$ being a power of 2. $\neg(ii)$
implies $\neg(iii)$. $(iv)$ within $\log$ follows from
\cite[Thm.3.9.1]{Li:97}.
$\neg(iv)$, since it does not even hold
within an additive constant (see \cite[p231]{Li:97}). $(v)$
immediate from definition. %
$\neg(vii)$ Define a
universal prefix Turing machine $U$ via some other universal
prefix Turing machine $U'$ by $U(00p)=U'(p)0$, $U(1p)=U'(p)1$,
$U(01)=\epstr$. For this $U$ we have $K(x0)=K(x1)+1\,\forall x$ ($K=K_U$),
which implies that $\Lambda_k$ for the error loss always predicts
$1$. %
$\neg(vi)$ follows from $\neg(vii)$.
\qed

Also, $K(x|\l(x))$ is a poor predictor, since $K(x0|\l(x0))\equa
K(x1|\l(x1))$, and the additive constant can be chosen to ones
need by an appropriate choice of $U$. Note that the larger a
semimeasure, the more distributions it dominates, the better its
predictive properties. This simple rule does not hold for
non-semimeasures. Although $M$ predicts better than $m$ predicts
better than $k$ in accordance with (\ref{Krels}),
$2^{-K(x|\l(x))}\geqm\MM(x)$ is a bad predictor disaccording with
(\ref{Krels}).

\paradot{Simple MDL}
There are other ways than $m$ of using shortest programs for
predictions. We have chosen the (in our opinion) most natural and
promising way.
A somewhat simpler version of MDL is to take the shortest
(nonhalting) program $p$ which outputs $x$, continue running $p$,
and use the continuation $y$ of $x$ for prediction:
\ifjcss
\bqan
  \widetilde m_|(x_t|x_{<t}) &:=& 1 \mbox{ if shortest program for } x_{<t}\!*
    \mbox{ computes } x_{<t}x_t*,
\\
  \widetilde m_|(\bar x_t|x_{<t}) &:=& 0.
\eqan
\else
\beqn
  \mbox{$\widetilde m_|(x_t|x_{<t}):=1$ if shortest program for $x_{<t}*$ computes
   $x_{<t}x_t*$,
   $\qquad\widetilde m_|(\bar x_t|x_{<t}):=0$.}
\eeqn
\fi
\ftheorem{thmsProp}{Properties of $\widetilde m$}{
For the simple MDL predictor $\widetilde m_|(x_t|x_{<t})$ and
$\widetilde m(x_{1:n}):=\prod_{t=1}^n\widetilde m_|(x_t|x_{<t})$, the
following holds: $\widetilde m$ is a deterministic, $(ii)$
monotone, $(iii)$ measure, satisfying $(iv)$ \mrcp\ (by
definition), is $\neg(i)$ not universal w.r.t.\
$\M_{comp}^{msr}\cap\M_{det}$, and is $\neg(v)$ not enumerable, and
is $\neg(vi)$ not convergent and $\neg(vii)$ not self-optimizing
w.r.t.\ some $\mu\in\M_{comp}^{msr}$.
}

Note that $\widetilde m_|$ contains more information than
$\widetilde m$. $\widetilde m_|$ cannot be reconstructed from
$\widetilde m$, since $\widetilde m_|(x_t'|x_{<t})$ is defined
even if $\widetilde m(x_{<t})=0$. $\neg(vi)$ and $\neg(vii)$
follow from non-denseness $\{\widetilde m_|\}=\B$. For $\neg(i)$
take $\omega=1^\infty$ in case $\widetilde m(1)=0$, and $0^\infty$
otherwise. We did not check the convergence properties for
deterministic environments.

Another possibility is to define
$m=f(\Km)$ with $f$ some monotone decreasing function other than
$f(\Km)=2^{-\Km}$, since $m=2^{-\Km}$ is not a semimeasure anyway.
We do not expect exciting results.

\section{Outlook and Open Problems}\label{secOpen}

\paradot{Speed of off-sequence convergence of $m$ for computable environments}
A more detailed analysis of the speed of convergence of $m(\bar
x_t|x_{<t})$ to zero in deterministic environments would be
interesting: How close are the off-sequence upper bound $(vi_4)$
$\eqm\Km^3$ and the lower bound $\neg(vi_5)$ $2^s-2$. Can the
lower bound be improved to $2^s\cdot\Km$? Maybe for the witnesses
of $m\not\eqm M$? The upper bound can be improved to $\eqm
\Km^2\cdot\log\Km$. Can the bound be improved to $\eqm\Km$?
Probably the most interesting open question is whether there exist
universal Turing machines for which the multiplicative constant is
of reasonable size. We expect that these hypothetical TMs, if they
exist, are very natural in the sense that they also possess other
convenient properties.

\paradot{Non-self-optimization for general $U$ and $\ell$}
Another open problem is whether for every non-degenerate
loss-function, self-optimization of $\Lambda_m$ can be violated.
We have shown that this is the case for particular choices of
the universal Turing machine $U$. If $\Lambda_m$
were self-optimizing for some $U$ and general loss, this would be an
unusual situation in Algorithmic Information Theory, where
properties typically hold for all or no $U$.
So we expect $\Lambda_m$ not to be self-optimizing for general
loss and $U$ (particular $\mu$ of course).
A first step may be to try to prove that for all $U$ there exists
a computable sequence $x_{1:\infty}$ such that $K(x_{<t}\bar
x_t)<K(x_{<t}x_t)$ for (infinitely) many $t$
(which shows $\neg(vii)$ for $K$ and error loss), and then try to
generalize to probabilistic $\mu$, $\Km$, and general loss
functions.

\paradot{Other complexity measures}
This work analyzed the predictive properties of the monotone
complexity $\Km$. This choice was motivated by the fact that $m$
is the MDL approximation of the sum $M$, and $\Km$ is {\em very}
close to $\KM$. We expect all other (reasonable) alternative
complexity measure to perform worse than $\Km$. But we should be
careful with precipitative conclusions, since closeness of
unconditional predictive functions not necessarily implies good
prediction performance, so distantness may not necessarily imply
poor performance. Besides the discussed
prefix Kolmogorov complexity $K$ \cite{Levin:74,Gacs:74,Chaitin:75}, %
monotone complexity $\Km$ \cite{Levin:73random}, and %
\SL's universal prior $M=2^{-\KM}$ \cite{Solomonoff:64,Solomonoff:78,Zvonkin:70}, %
one may investigate the predictive properties of the %
plain Kolmogorov complexity $C$ \cite{Kolmogorov:65},
process complexity \cite{Schnorr:73}, %
Chaitin's complexity $K\!c$ \cite{Chaitin:75}, %
extension semimeasure $M\!c$ \cite{Cover:74}, %
uniform complexity \cite{Loveland:69ic,Loveland:69acm}, %
cumulative $K^E$ and general $K^G$ complexity and corresponding measures \cite{Schmidhuber:02gtm}, %
predictive complexity $K\!P$ \cite{Vovk:98}, %
speed prior $S$ \cite{Schmidhuber:02speed}, %
Levin complexity \cite{Levin:73search,Levin:84}, %
and several others.
Most of them are described in \cite{Li:97}.
Many properties and relations are known for the unconditional
versions, but little relevant for prediction of the conditional
versions is known.

\paradot{Two-part MDL}
We have approximated $M(x):=\sum_{p:U(p)=x*}2^{-\l(p)}$ by its
dominant contribution $m(x)=2^{-\Km(x)}$, which we have
interpreted as deterministic or one-part universal MDL. There is
another representation of $M$ due to Levin \cite{Zvonkin:70} as a
mixture over semimeasures:
$M(x)=\sum_{\nu\in\M_{enum}^{semi}}2^{-K(\nu)}\nu(x)$ with
dominant contribution $m_2(x)=2^{-\Km_2(x)}$ and universal
two-part MDL
$\Km_2(x):=\min_{\nu\in\M_{enum}^{semi}}\{-\lb\,\nu(x)+K(\nu)\}$.
MDL ``lives'' from the validity of this approximation.
$K(\nu)$ is the complexity of the probabilistic model $\nu$, and
$-\lb\,\nu(x)$ is the (Shannon-Fano) description length of data
$x$ in model $\nu$.
MDL usually refers to two-part MDL, and not to one-part MDL. A
natural question is to ask about the predictive properties of
$m_2$, similarly to $m$. $m_2$ is even closer to $M$ than $m$ is
($m_2\eqm M$), but is also not a semimeasure. Drawing the analogy
to $m$ further, one may ask whether (slow) posterior convergence
$m_2\to\mu$ w.p.1 for computable probabilistic environments $\mu$
holds. In \cite{Hutter:04mdl2p,Hutter:04mdlspeed} we show, more
generally, slow posterior convergence of two-part MDL w.p.1 in
probabilistic environments $\mu$. See also \cite{Barron:91}, for
convergence results for two-part MDL in i.i.d.\ environments.

\paranodot{More abstract proofs}
showing that violation of some of the criteria $(i)-(iv)$
necessarily lead to violation of $(vi)$ or $(vii)$ may deal with a
number of complexity measures simultaneously. For instance, we
have seen that any non-dense posterior set $\{\tilde
k(x_t|x_{<t})\}$ implies non-convergence and
non-self-optimization in probabilistic environments; the
particular structure of $m$ did not matter.
Maybe a probabilistic version of Theorem~\ref{thCUNSM} on the
convergence of universal non-semimeasures is possible under some
(mild?) extra assumptions on $b$.

\paradot{Extra conditions}
Non-convergence or non-self-optimization of $m$ do not
necessarily mean that $m$ fails in practice. Often one knows more
than that the environment is (probabilistically) computable, or
the environment possess certain additional properties, even if
unknown. So one should find sufficient and/or necessary extra
conditions on $\mu$ under which $m$ converges / $\Lambda_m$
self-optimizes rapidly. The results of this work have shown that
for $m$-based prediction one {\em has} to make extra assumptions
(as compared to $\MM$). It would be interesting to characterize
the class of environments for which universal MDL alias $m$ is a
good predictive approximation to $M$. Deterministic computable
environments were such a class, but a rather small one, and
convergence can be slow.

\addcontentsline{toc}{section}{References}
\begin{small}

\end{small}

\end{document}